\documentclass[]{siamltex} 
\usepackage{latexsym,url,epsfig}
\usepackage[USenglish]{babel}
\usepackage{enumerate}
\usepackage{umlaut}
\usepackage{psfrag}
\usepackage{url}
\usepackage{booktabs}

\newcommand{\uncomment}[1]{{}}

\title{
Higher-Dimensional Packing with Order Constraints
\thanks{Preliminary extended abstract versions reporting on parts of this paper
appeared in \cite{extending,fkt-mdpoc-01}.}
}

\author{
S\'{a}ndor P.\ Fekete\thanks{
Department of Mathematical Optimization,
Braunschweig University of Technology,
D--38116 Braunschweig, Germany,
{\tt s.fekete@tu-bs.de}. 
Partially supported by 
Deutsche Forschungsgemeinschaft (DFG) within the special
focus program ``Reconfigurable Computing'' (SPP 1148).
}
\and
Ekkehard K\"ohler\thanks{
Department of Mathematics,
TU Berlin,
D--10623 Berlin, Germany,
{\tt ekoehler@math.tu-berlin .de}.}
\and
J{\"u}rgen Teich\thanks{
Department of Computer Science 12,
(Hardware-Software-Co-Design),
University of Erlangen-Nuremberg,
D-91058 Erlangen,
Germany,
{\tt teich@informatik.uni-erlangen.de}.
Partially supported by 
Deutsche Forschungsgemeinschaft (DFG) within the special
focus program ``Reconfigurable Computing'' (SPP 1148).
}
}

\begin{document}

\maketitle

\begin{abstract}
  We present a first exact study on higher-dimensional packing
  problems with order constraints. Problems of this type occur
  naturally in applications such as logistics or computer architecture
  and can be interpreted as higher-dimensional generalizations of
  scheduling problems.  Using graph-theoretic structures to describe
  feasible solutions, we develop a novel exact branch-and-bound
  algorithm.  This extends previous work by Fekete and Schepers; a key
  tool is a new order-theoretic characterization of feasible
  extensions of a partial order to a given complementarity graph that
  is tailor-made for use in a branch-and-bound environment.  The
  usefulness of our approach is validated by computational results.
\end{abstract}



\pagestyle{myheadings}
\thispagestyle{plain}
\markboth{S. P. FEKETE, E. K\"OHLER, AND J. TEICH}{HIGHER-DIMENSIONAL PACKING WITH ORDER CONSTRAINTS}

\section{Introduction}\label{sec:intro}

\paragraph{Scheduling and Packing Problems}

Scheduling is arguably one of the most
important topics in combinatorial optimization. Typically, we are
dealing with a one-dimensional set of objects (``jobs'') that need to
be assigned to a finite set of containers (``machines'').  Problems of
this type can also be interpreted as (one-dimensional) packing
problems, and they are NP-hard in the strong sense, as problems like
3-{\sc Partition} are special cases.

Starting from this basic scenario, there are different generalizations
that have been studied. Many scheduling problems have \emph{precedence
  constraints} on the sequence of jobs.  On the other hand, a great
deal of practical packing problems consider \emph{higher-dimensional}
instances, where objects are axis-aligned boxes instead of intervals.
Higher-dimensional packing problems arise in many industries, where
steel, glass, wood, or textile materials are cut.  The
three-dimensional problem is important for practical applications such
as container loading.

In this paper, we give the first study of problems that comprise both
generalizations: these are higher-dimensional packing problems with
order con\-straints---or, from a slightly different point of view,
higher-dimensio\-nal sche\-du\-ling problems.  In higher-dimensional
packing, these prob\-lems arise when dealing with precedence
constraints that are present in many container-loading problems.
Another practical motivation for considering multi-dimensional
scheduling problems arises from optimizing the reconfiguration of a
particular type of computer chips called FPGAs---described below.

\paragraph{FPGAs and Higher-Dimensional Scheduling}

\begin{figure}[htbp]
  \begin{center}
  \psfrag{t}{$t$}
  \psfrag{x}{$x$}
  \psfrag{y}{$y$}
  \includegraphics[width=9cm]{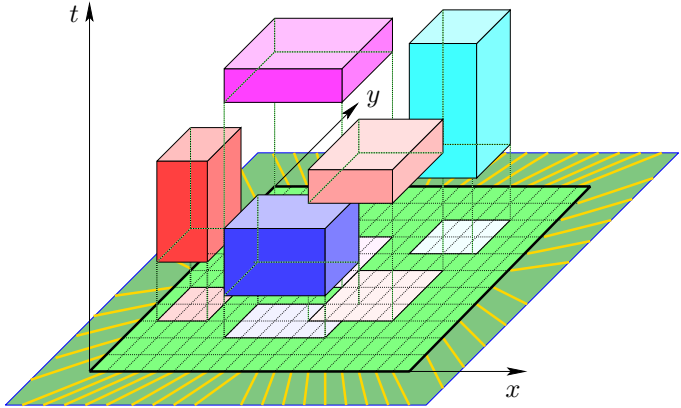}
  \caption{An FPGA and a set of five jobs, shown as
    projections in ordinary two-dimensional space and in
    three-dimensional space-time.  Jobs must be placed inside the chip
    and must not overlap if executed simultaneously on the chip.}
  \label{fig:fpga}
  \end{center}
\end{figure}

A particularly interesting class of instances of three-dimensional
orthogonal packing arises from a new type of reconfigurable computer
chips, called \emph{field-programmable gate arrays} (FPGAs).  An FPGA
typically con\-sists of a regular rectangular grid of equal
configurable cells (logic blocks) that allow the prototyping of simple
logic functions together with simple registers and with special
routing resources (see Figure~\ref{fig:fpga}).  These chips (see e.g.
\cite{Atm,Xil}) may support several independent or interdependent jobs
and designs at a time, and parts of the chip can be reconfigured
quickly during run-time.  (For more technical details on the
underlying architecture, see the previous
paper~\cite{TFS00}, and the more recent abstract~\cite{DATE2001}.)
Thus, we are faced with a general class of problems that can be seen
both as scheduling and packing problems.  In this paper, we develop a
set of mathematical tools to deal with these \emph{higher-dimensional
  scheduling problems}, and we show that our methods are suitable for
solving instances of interesting size to optimality.

\paragraph{Related Work}

We are not aware of any exact study of higher-dimensional packing or
sche\-duling problems with order constraints.  For a comprehensive survey of
classical ``one-dimensional'' scheduling problems, the reader is
referred to \cite{sched}. A related problem 
is dynamic storage allocation, where ``processing jobs'' means
storing them in contiguous blocks of memory from a one-dimensional array.
Considering time as the second dimension leads to a two-dimensional
packing problem, possibly with order constraints. However, this
problem is primarily an online problem; for example, see \cite{lno-tbdsa-94}.
In an offline setting, precise starting an ending time values 
imply order constraints, but also provide more information.
(See our paper \cite{TFS00} for exact methods for that scenario.)

Closest to our problems is the class of
\emph{resource-constrained project scheduling problems}
(RCPSP), which can be interpreted as a step towards higher-dimensional
packing problems: In addition to a duration $t_i$ and precedence
constraints on the temporal order of jobs, each job $i$ may have a number
of other ``sizes'' $x^{(1)}_i,\ldots,x^{(k)}_i$; $x^{(j)}_i$ 
indicates the amount of resource $j$ required for the proessing of
job $i$.  The total amount $\sum_i x^{(j)}_i$ of each resource $j$ is
limited at any given time. See the book \cite{Weg} and the references
in the article \cite{MSSU} for an extensive survey of recent work in
this area.  Even though RCPSPs can be formulated as integer problems,
solving resource-constrained scheduling problems is already quite hard
for instances of relatively moderate size: The standard benchmark
library used in this area consists of instances with 30, 60, 90 and
120 jobs. Virtually all work deals with lower and upper bounds on
these instances, and even for instances with 60 jobs, a considerable
number has not yet been solved to optimality.

\begin{figure}[htbp]
  \begin{center}
\hfill
  \includegraphics[width=3.5cm]{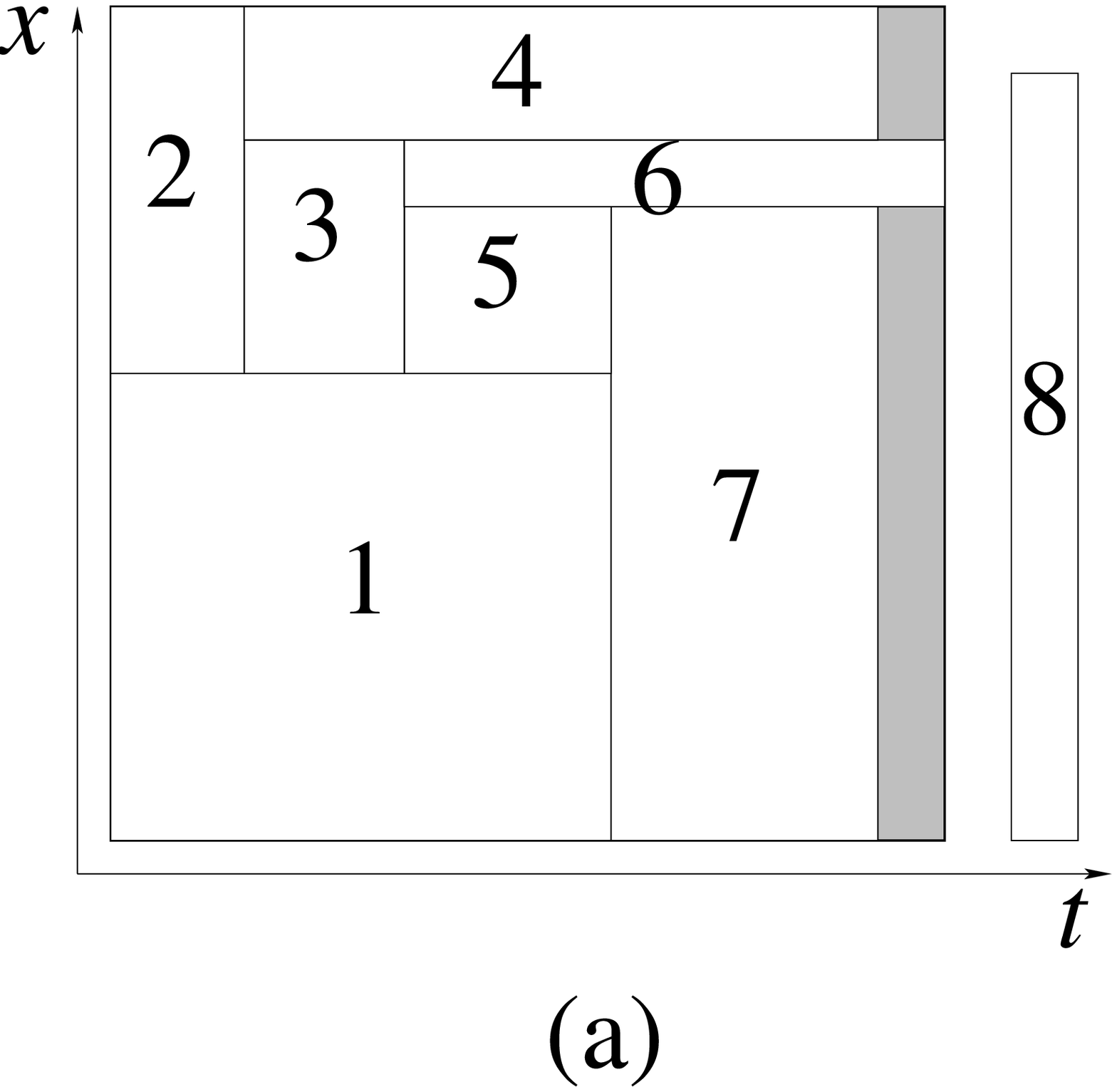}
\hfill
  \includegraphics[width=3.5cm]{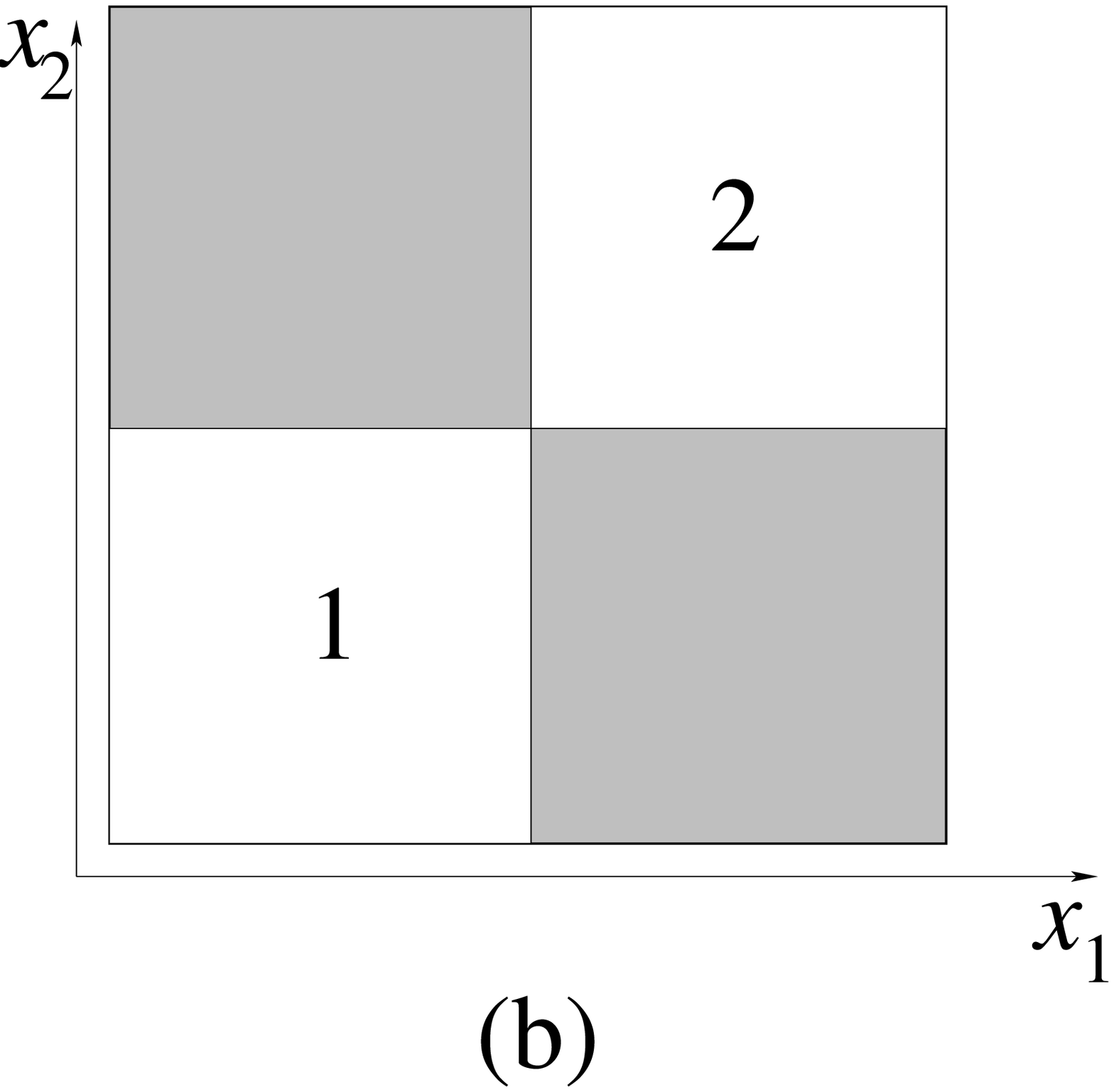}
\hfill
  \includegraphics[width=3.5cm]{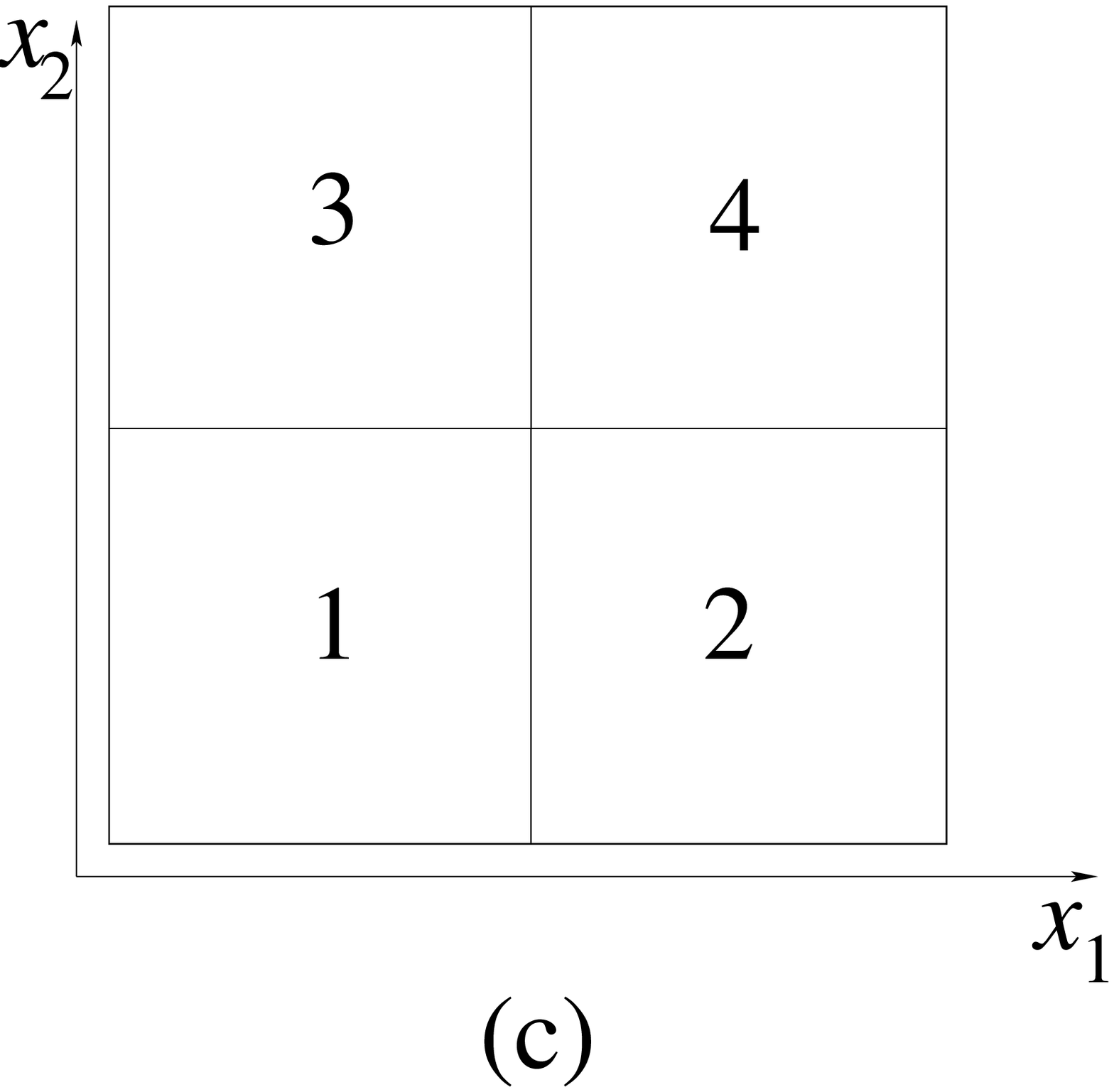}
\hfill
  \caption{Differences between RCPSP and packing:
(a) A set of jobs that is feasible for
    scheduling with one resource constraint, but infeasible for
    two-dimensional packing: Job 8 does not violate a resource
    constraint, but does not fit as a contiguous rectangle.
(b) A set of jobs that is just feasible for RCPSP with $d=2$ constraints, i.e.,
that does not allow any tighten of either constraint without becoming infeasible. 
(c) A set of boxes that is just feasible for packing in $d=2$ dimensions.}
  \label{fig:nosched}
  \end{center}
\end{figure}

It is easy to see that two-dimensional packing problems
(possibly with precedence constraints on the temporal order) can be
relaxed to a scheduling problem with one resource-constraint,  by allowing 
a non-contiguous use of resources, i.e., the higher-dimensional 
analogue of preemption. However, the
example in Figure~\ref{fig:nosched} shows that the converse is not
true, even for small instances of two-dimensional packing problems
without any precedence constraints: An optimal solution for the
corresponding resource-constrained scheduling problem may not
correspond to a feasible arrangement of rectangles for the original
packing problem.  (We leave it to the reader to verify the latter
claim.) For $d\geq 2$ the difference becomes more pronounced:
The $d$ knapsack constraints for RSPSP require that for 
{\em all} of the $d$ individual resources and 
every pair of jobs, a disjointness
property must be satisfied; on the other hand,
the more geometric conditions on $d$-dimensional packing 
require that any pair of boxes must be disjoint
in {\em at least one} of their coordinate intervals.
Arguably, the disjunctive constraints on $(d+1)$-dimensional 
packing problems are harder to model.

Higher-dimensional packing problems (without order constraints) have
been considered by a great number of authors, but only few of them
have dealt with the exact solution of general two-dimensional
problems.  See \cite{ESA,fs-cchdop-04} for an overview.  It should be
stressed that unlike one-dimensional packing problems,
higher-dimensional packing problems allow no straightforward
formulation as integer programs: After placing one box in a container,
the remaining feasible space will in general not be convex. Moreover,
checking whether a given set of boxes fits into a particular container
(the so-called \emph{orthogonal packing problem}, {\bf OPP}) is trivial in
one-dimensional space, but NP-hard in higher dimensions.

Nevertheless, attempts have been made to use standard approaches of
mathematical programming.  Beas\-ley~\cite{Bea85} and
Hadjiconstantinou and Chris\-to\-fides~\cite{HC95} have used a
discretization of the available positions to an underlying grid to get
a 0-1 program with a pseudopolynomial number of variables and
constraints. Not surprisingly, this approach becomes impractical
beyond instances of rather moderate size.  More recently,
Padberg~\cite{Pad00} gave a \emph{mixed integer programming}
formulation for three-dimensional packing problems, similar to the one
anticipated by Schepers~\cite{Sch97} in his thesis.  Padberg expressed
the hope that using a number of techniques from branch-and-cut will be
useful; however, he did not provide any practical results to support
this hope.

In~\cite{fs-eahdop-06,ESA,fs-cchdop-04,fs-gfbhdop-04,TFS00}, 
a different approach to
characterizing feasible packings and constructing optimal solutions is
described.  A graph-theoretic characterization of the relative
position of the boxes in a feasible packing (by so-called
\emph{packing classes}) is used, representing $d$-dimensional packings
by a $d$-tuple of interval graphs (called \emph{component graphs})
that satisfy two extra conditions. This factors out a great deal of
symmetries between different feasible packings, it allows to make use
of a number of elegant graph-theoretic tools, and it reduces the
geometric problem to a purely combinatorial one without using
brute-force methods like introducing an underlying coordinate grid.
Combined with good heuristics for dismissing infeasible sets of
boxes~\cite{IPCO}, a tree search for constructing feasible packings
was developed. This exact algorithm has been implemented; it
outperforms previous methods by a clear margin.

For the benefit of the reader, a concise description of this approach
is contained in Section~\ref{sec:opp}.

\paragraph{Graph Theory of Order Constraints}

In the context of scheduling with precedence constraints, a natural
problem is the following, called \emph{transitive ordering with
  precedence constraints} ({\bf TOP}):
  Consider a partial order $P=(V,\prec)$ of precedence
constraints and a (temporal) comparability graph $G=(V,E)$, such that
all relations in $P$ are represented by edges in $G$.  Is there a
transitive orientation $D=(V,A)$ of $G$, such that $P$ is contained in
$D$?

Korte and M\"ohring \cite{KM} have given a linear-time algorithm for
deciding {\bf TOP}, using modified PQ-trees. However, their approach
requires knowledge of the full set of edges in $G$. When running a
branch-and-bound algorithm for solving a scheduling problem, these
edges of $G$ are only known partially during most of the tree search,
but already this partial edge-set may
prohibit the existence of a feasible solution for a given partial
order $P$.  This makes it desirable to come up with structural
characterizations that are already useful when only parts of $G$ are
known.

\begin{figure}[htbp]
  \begin{center}
    \includegraphics[width=8.0cm]{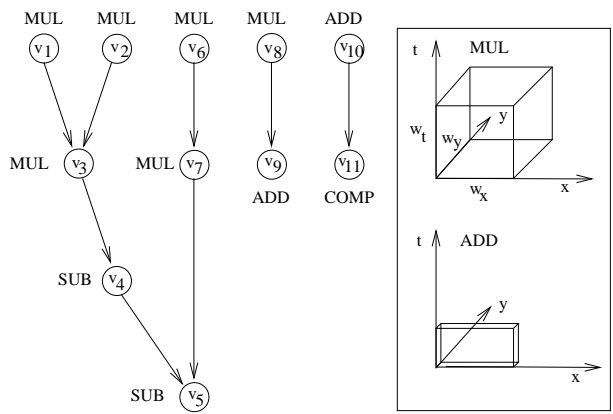}
    \caption{Dependency graph of jobs and shape of
      modules (3D boxes) with the spatial dimensions \boldmath{$x$}
      and \boldmath{$y$} and the temporal dimension \boldmath{$t$}
      (execution time).}
    \label{fig:pdedep}
  \end{center}
\end{figure}

Such a set of precedence constraints may be described by a dependency
graph, see Figure~\ref{fig:pdedep}.

For a problem instance of this type, we describe a general framework
for finding exact solutions to the problem of minimizing the height of
a container of given base area, or minimizing the makespan of a
higher-dimensional non-preemptive scheduling problem. 

\paragraph{Results of this paper}

In this paper, we give the first exact study of higher-dimensional
packing with order constraints, which can also be interpreted as
\emph{higher-dimensional non-preemptive
  scheduling problems}.  We develop a general framework for problems
of this type by giving a pair of necessary and sufficient conditions
for the existence of a solution for the problem {\bf TOP} on graphs $G$ in
terms of forbidden substructures.  Using the concept of packing
classes, our conditions can be used quite effectively in the context
of a branch-and-bound framework, because it can recognize infeasible
subtrees at ``high'' branches of the search tree.  In particular, we
describe how to find an exact solution to the problem of minimizing
the height of a container of given base area. If this third dimension
represents time, this amounts to minimizing the makespan of a
higher-dimensional scheduling problem.  We validate the usefulness of
these concepts and results by providing computational results. Other
problem versions (like higher-dimensional knapsack or bin packing
problems with order constraints) can be treated similarly.

The rest of this paper is organized as follows.  In
Section~\ref{sec:prelim}, we describe basic assumptions and some
terminology.  The notion of packing classes and a solution to packing
problems without precedence constraints is summarized in
Section~\ref{sec:opp}.  In Section~\ref{sec:copp}, we introduce
precedence constraints, describe the mathematical foundations for
incorporating them into the search, and explain how to implement the
resulting algorithms.  Section~\ref{sec:impl-class-modul} provides the
necessary mathematical foundations for the correctness of our
approach.  Finally, we present computational results for a number of
different benchmarks in Section~\ref{sec:exp}.

\medskip
\section{Preliminaries}
\label{sec:prelim}

An FPGA consists of a rectangular grid of identical logic cells.
Each job $v$ (or ``module'') requires a rectangle 
of size $w_x(v)$ by $w_y(v)$ with fixed axis-parallel orientation,
and needs to remain available for at least the time
$w_t(v)$. Any logic cell that is not occupied by a module
may be used by one of the rectangular jobs. As shown in Figure~\ref{fig:fpga},
we are dealing
with a three-dimensional packing problem, possibly with 
order constraints. In the following, we describe
technical as well as mathematical terminology and assumptions.

\medskip
\subsection{Architecture Assumptions}

The model of having relocatable, rectangular modules
is justified by current FPGA technology \cite{Atm,Xil}.

\paragraph{Intermodule communication}

Intermodule communication is assumed to occur at the end of operation
of the sending module (task model).  The issuing module may store its
result register values into an external memory connected to the FPGA
interface (read-out) via a bus interface. Memory is allocated for temporary 
storage of intermediate results\footnote{A static memory allocation
  may be deduced directly from the static placement.}.  Afterwards, the
receiving module will read the communicated data into its registers
via the bus interface.  With this communication style, it is
justifiable to ignore routing overhead between modules that otherwise
might introduce additional placement constraints.

\paragraph{I/O-overhead}

The communication time needed for writing out and reading in
communicated data may be accounted for by considering this as an
offset and being part of the execution time of a job.

\paragraph{Reconfiguration overhead}

The time needed for carrying out reconfigurations may be modeled by a
constant (possibly a different number for each job), depending on the
target architecture.  This may be considered a simplification because
the reconfiguration time might depend on the result of the placement.
Consider two equal modules with identical placements.  A
reconfiguration for the second module might not be necessary in case
no third module is occupying a (sub)set of cells in the time interval
between the execution of the two modules.  However, there are
many different models for accounting for  reconfiguration times,
and the particular choice should be adapted individually to the 
target architecture.

\medskip
\subsection{Mathematical Terminology}
\label{sec:math-term}

\paragraph{Problem instances}

We assume that a problem instance is given by a \emph{set $V$ of
  jobs}.  Each job has a \emph{spatial requirement} in the $x$- and
$y$-direction, denoted by $w_x(v)$ and $w_y(v)$, and a
\emph{duration}, denoted by a size $w_t(v)$ along the time axis.  The
available space $H$ consists of an area of size $h_x\times h_y$.  In
addition, there may be an overall allowable time $h_t$ for all jobs to
be completed.  A \emph{schedule} is given by a start time $p_t(v)$ for
each job. A schedule is \emph{feasible}, if all jobs can be carried
without preemption or overlap of computation jobs in time and space, such that
all jobs are within spatial and temporal bounds.

\paragraph{Graphs}

Some of our descriptions make use of a number of certain
graph-theoretic concepts. An (undirected) graph $G=(V,E)$ is given by
a set of vertices $V$, and a set of edges $E$; each edge describes the
adjacency of a pair of vertices, and we write $\{u,w\}$ for an edge
between vertices $u$ and $w$.  We only consider graphs without
multiple edges and without loops. For a graph $G$, we obtain the
\emph{complement graph} $\overline{G}$ by exchanging the set $E$ of
edges with the set $\overline{E}$ of non-edges.  In a directed graph
$D=(V,A)$, edges are oriented, and we write $(u,w)$ to denote an edge
directed from $u$ to $w$.  A graph $G=(V,E)$ is a \emph{comparability
  graph} 
if there is a \emph{transitive orientation} for it,
  i.e., the edges $E$ can be oriented to a set of directed arcs $A$,
  such that we get the transitive closure of a partial order.  More
  precisely, this means that $D=(V, A)$ is a cycle-free digraph for
  which the existence of edges $(u,v)\in A$ and $(v,w)\in A$ for any
  $u,v,w\in V$ implies the existence of $(u,w)\in A$.  
Comparability graphs have a variety of nice
  properties.  For our purpose we will make use of the algorithmic
  result that computing maximum weighted cliques on comparability
  graphs can be done efficiently (see \cite{GOL80}).  
A
  closely related family of graphs, the \emph{interval graphs}, are
  defined as follows.  Given a set of intervals on the real line,
  every vertex of the graph corresponds to
  an interval of the set; two vertices are joined by an edge if the
  corresponding intervals have a non-empty intersection.  Interval
  graphs have been studied intensively in graph theory (see
  \cite{GOL80,RHM}), and, similar to comparability graphs, they have
  a number of very useful algorithmic properties.

\paragraph{Precedence constraints}

Mathematically, a set of precedence constraints is given by a partial
order $P=(V,\prec)$ on $V$.  The relations in $\prec$ can be
interpreted as a directed acyclic graph
$D_P=(V,A_P)$, where $A_P$ is a set of directed arcs corresponding to
the relations in $\prec$. In the presence of such a
partial order, a feasible schedule is required to
satisfy the capacity constraints of the container, as well as these
additional constraints.

\paragraph{Packing problems}

In the following, we treat jobs as axis-aligned $d$-dimensional
boxes with given orientation, and feasible schedules as arrangements
of boxes that satisfy all side constraints. This
is implied by the term of a \emph{feasible packing}.  There may be
different types of objective functions, corresponding to different
types of packing problems. The \emph{Orthogonal Packing Problem} ({\bf OPP})
is to decide whether a given set of boxes can be placed within a given
``container'' of size $h_x\times h_y\times h_t$.  For the Constrained
OPP ({\bf COPP}), we also have to satisfy a partial order $P=(V,\prec)$ of
precedence constraints in the $t$-dimension.  To emphasize the
motivation of temporal precedence constraints, we write $t$ to suggest
that the time coordinate is constrained, and $x$ and $y$ to imply that
the space coordinates are unrestricted.  Although our application
mainly requires to consider those temporal constraints, it should be
mentioned that our approach works the same way when
dealing with spatial restrictions; that is why we are using
a generic index $i$ in the mathematical discussion, while
some of our benchmark examples consider a temporal dimension $t$.

There are various optimization problems that have {\bf OPP} or {\bf COPP} as their
underlying decision problems.  The \emph{Base Minimization Problem}
(BMP) is to minimize the size $h_x$ for a fixed $h_t$ such that all
boxes fit into a container $h_x\times h_x\times h_t$ with quadratic
base.  This corresponds to minimizing the necessary area to carry out
a set of computations within a given time.  Because our main
motivation arises from dynamic chip reconfigurations, where we want to
minimize the overall running time, we focus on the \emph{Constrained
  Strip Packing Problem} (CSPP), which is to minimize the size $h_t$
for a given base size $h_x\times h_y$, such that all boxes fit into
the container $h_x\times h_y\times h_t$.  Clearly, we can use a
similar approach for other objective functions.

\medskip
\section{Solving Unconstrained Orthogonal Packing Problems}
\label{sec:opp}
\subsection{A General Framework}

If we have an efficient method for solving {\bf OPPs}, we can also solve
BMPs and SPPs by using a binary search.  However, deciding the
existence of a feasible   packing
is a hard problem in higher dimensions, and methods proposed 
by other authors \cite{Bea85,HC95} have been of limited success.

Our framework uses a combination of different approaches to overcome
these problems:
\begin{enumerate}
\item Try to disprove the existence of a packing by 
    classes of lower bounds on the necessary size.
\item In case of failure, try to find a feasible packing by using fast
  heuristics.
\item If the existence of a packing is still unsettled, start an
  enumeration scheme in form of a branch-and-bound tree search.
\end{enumerate}

By developing good new bounds for the first stage, we have been able
to achieve a considerable reduction of the number of cases in which a
tree search needs to be performed.  (Mathematical details for this
step are described in \cite{IPCO,fs-gfbhdop-04}.)  However, it is clear that
the efficiency of the third stage is crucial for the overall running
time when considering difficult problems. Using a purely geometric
enumeration scheme for this step by trying to build a partial
arrangement of boxes is easily seen to be immensely time-consuming. In
the following, we describe a purely combinatorial characterization of
feasible packings that allows to perform this step more efficiently.

\medskip
\subsection{Packing Classes}
\label{sec:packing-classes}

\begin{figure}[htbp]
  \begin{center}
    \includegraphics[width=10cm]{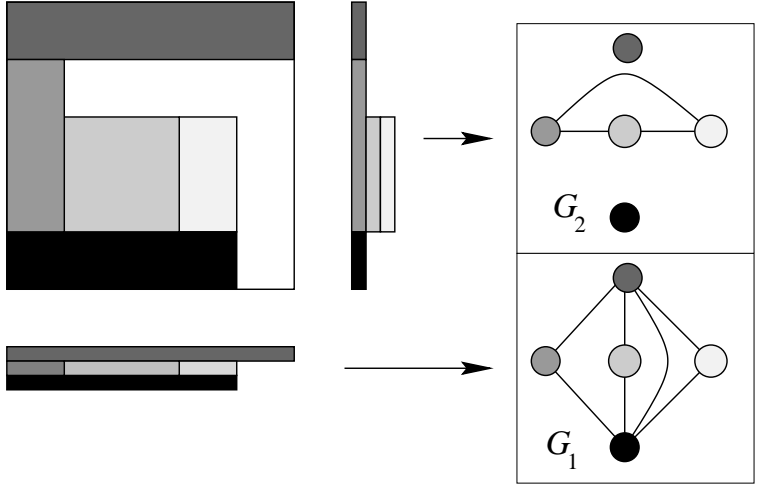}
    \caption{The projections of the boxes onto the
      coordinate axes define interval graphs (here in 2D:
      \boldmath{$G_1$} and \boldmath{$G_2$}).}
    \label{fig:intervgra}
  \end{center}
\end{figure}

Consider a feasible packing in $d$-dimensional space, and project the
boxes onto the $d$ 
  coordinate axes.  This converts the one
$d$-dimensional arrangement into $d$ one-dimensional ones (see
Figure~\ref{fig:intervgra} for an example in $d=2$).  By disregarding
the exact coordinates of the resulting intervals in direction $i$ and
only considering their intersection properties,
  we get the \emph{component graph}
$G_i=(V,E_i)$: Two boxes $u$ and $v$ are connected by an edge in
$G_i$, iff their projected intervals in direction $x_i$ have a
non-empty intersection.
By definition, these graphs are
\emph{interval graphs}. 

Considering sets of $d$ component graphs $G_i$ instead of complicated
geometric arrangements has some clear advantages (algorithmic
implications for our specific purposes are discussed further down).
It is not hard to check that the following three conditions must be
satisfied by all $d$-tuples of graphs $G_i$ that are constructed from
a feasible packing:

\bigskip
\begin{enumerate}[\bf C1:]
\item $G_i$ is an interval graph, $\forall i \in \{1,\cdots,d\}$.
\item Any independent set $S$ of $G_i$ is $i$-admissible, $\forall i
  \in \{1,\cdots,d\}$, i.e., $w_i(S) = \sum_{v \in S} w_i(v) \leq
  h_i$, because all boxes in $S$ must fit into the container in the
  $i$th dimension.
\item $\cap_{i=1}^{d} E_i = \emptyset$. In other words, there must be
  at least one dimension in which the corresponding boxes do not
  overlap.
\end{enumerate}

A $d$-tuple of component graphs satisfying these necessary conditions
is called a \emph{packing class}.  The remarkable property (proven in
\cite{Sch97,fs-cchdop-04})
  is that these three conditions are also sufficient for the
existence of a feasible packing.

\medskip
\begin{theorem}[Fekete, Schepers]
  A set of $d$-dimensional boxes allows a feasible packing,
  iff there is a packing class, i.e., a $d$-tuple of graphs $G_i=(V,E_i)$ 
  that satisfies the conditions {\bf C1}, {\bf C2}, {\bf C3}.
\end{theorem}

\medskip
This allows us to consider only packing
classes in order to decide the existence of a feasible packing, and to
disregard most of the geometric information.

\subsection{Solving OPPs}
\label{sec:solving-opps}

Our search procedure works on packing classes, i.e., $d$-tuples of
component graphs with the properties {\bf C1}, {\bf C2}, {\bf C3}.  Because each packing
class represents not only a single packing but a whole family of
equivalent packings, we are effectively dealing with more than one
possible candidate for an optimal packing at a time.  (The reader may
check for the example in Figure~\ref{fig:intervgra} that there are 36
different feasible packings that correspond to the same packing
class.)

For finding an optimal packing, we use a branch-and-bound
approach. The search tree is traversed by
depth first search, see
\cite{fs-eahdop-06,Sch97} for details.  Branching is done by 
{deciding about a single pair of vertices $b,c$, whether the corresponding edge is contained in $E_i$ or is not contained in $E_i$, i.e.,}
$\{b,c\} \in E_{i}$ or $\{b,c\} \notin E_{i}$.  
{So in fact, there are three classes of edges; those which are fixed to be in $E_i$, those which are fixed not to be in $E_i$ (non-edges), and those for which it is not decided yet whether they will be contained in $E_i$ or not.}
After each branching
step, it is checked whether one of the three conditions {\bf C1}, {\bf C2}, {\bf C3} is
violated {with respect to the currently fixed edges and non-edges}; furthermore it is checked
whether a violation can only be avoided by fixing further {(formerly undecided)} edges or non-edges.
Testing for two of the conditions {\bf C1}--{\bf C3} is easy:
  enforcing {\bf C3} is obvious; 
  checking {\bf C2} can be done efficiently, since $\overline{G}_i$
    is a comparability graph and, as mentioned before, in those graphs
    maximum weighted cliques can be done efficiently.  {Note that for this step only non-edges are used, i.e., pairs of vertices for which has been decided already that they are not contained in $E_i$.}
    In order to ensure that property {\bf C1} is not
  violated, we use some graph-theoretic characterizations of interval
  graphs and comparability graphs.  These characterizations are based
  on two forbidden substructures. (Again, see \cite{GOL80} for details;
  the first condition is based on the classical characterizations by
  \cite{Gh62,GiHo64}: a graph is an interval graph \emph{iff} its
  complement has a transitive orientation, and it does not contain any
  induced chordless cycle of length 4.)  In particular, the following
  configurations have to be avoided:

\begin{enumerate}[\bf G1:]
\item induced chordless cycles of length 4 in $E_{i}$;
\item so-called 2-chordless odd cycles in the set of
  edges excluded from $E_i$ (see \cite{fs-eahdop-06,GOL80} for details);
\item infeasible stable sets in $E_i$.
\end{enumerate}
  Each time we detect such a fixed subgraph,
we can abandon the search on this node.  Furthermore, if we detect a
fixed subgraph, except for one unfixed edge, we can fix this edge,
such that the forbidden subgraph is avoided.

Our experience shows that in the considered examples
these conditions are already useful when only small subsets of edges
have been fixed, because by excluding small sub-configurations like
induced chordless cycles of length 4, each branching step triggers a
cascade of more fixed edges.

\section{Packing Problems with Precedence Constraints}
\label{sec:copp}

As mentioned in the above discussion, a key advantage of considering
packing classes is that it makes possible the consideration of packing problems
independent of precise geometric placement, and that it allows
arbitrary feasible interchanges of placements.
  However, for most practical instances, we have to satisfy
additional constraints for the temporal placement, i.e., for the
relative start times of jobs.
For our approach, the nature of the data structures may simplify these
problems from three-dimensional to purely two-dimensional ones: If the
whole schedule is given, all edges $E_t$ in one of the graphs are
determined, so we only need to construct the edge sets $E_x$ and $E_y$
of the other graphs.  As worked out in detail in~\cite{TFS99b,TFS00},
this allows it to solve the resulting problems quite efficiently if
the arrangement in time is already given.

A more realistic, but also more involved situation arises if only a
set of precedence constraints is given, but not the full schedule.  
We describe in the following how further mathematical tools in addition
to packing classes allow useful algorithms. Note that our method
of dealing with order constraints is not restricted to one
(the temporal) dimension; in fact, we can also deal with 
constraints in several dimensions at once, as demonstrated
in Section~\ref{sec:exp}, Figure~\ref{fig:sq}.

\subsection{Packing Classes and Interval Orders}
\label{sec:pack-class-interv}

Any edge $\{v_1,v_2\}$ in a component graph $G_i$ corresponds to an
intersection between the projections of
boxes $1$ and $2$ onto the $x_i$-axis.
This means that the complement graph $\overline{G_i}$ given by the
complement $\overline{E_i}$ of the edge set $E_i$ consists of all
pairs of coordinate intervals that are ``comparable'': Either the
first interval is ``to the left'' of the second, or vice versa. 

\begin{figure}[htbp]
  \begin{center}
    \includegraphics[width=10.5cm]{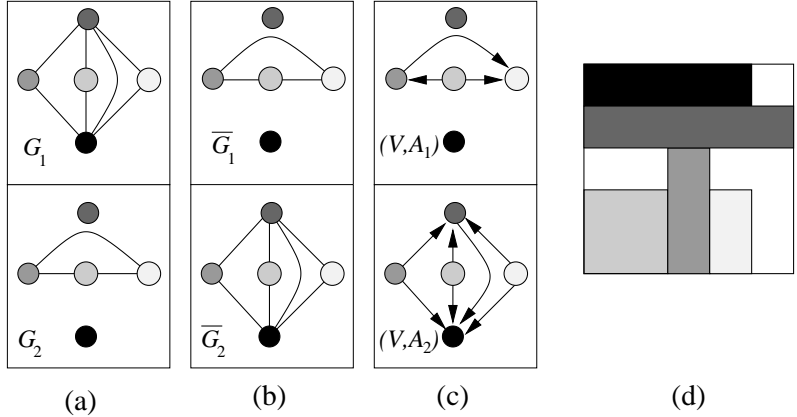}
    \caption{(a) A two-dimensional packing class.
      (b) The corresponding comparability graphs. (c) Two transitive
      orientations. (d) A feasible packing
      corresponding to the orientation.}
    \label{fig:orientation}
  \end{center}
\end{figure}

{Any (undirected) graph of this type is a comparability graph.
  By orienting edges to point from ``left'' to ``right'' intervals, we
  get a partial order of the set $V$ of vertices, a so-called
  \emph{interval order} \cite{Fishburn85,RHM}. Obviously, this order
  relation is transitive, inducing a transitive orientation on the
  (undirected) comparability graph $G_i$.  See
  Figure~\ref{fig:orientation} for a (two-dimensional) example of a
  packing class, the corresponding comparability graphs, the
  transitive orientations, and the packing corresponding to the
  transitive orientations.}

\begin{figure}[htbp]
  \begin{center}
    \includegraphics[width=9cm]{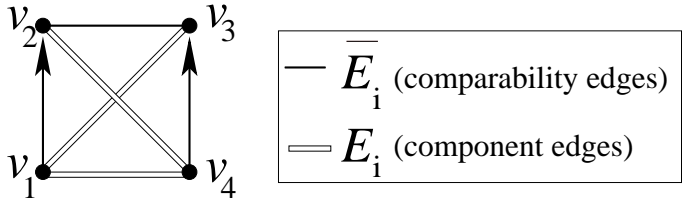}
    \caption{A comparability graph
      \boldmath{$\overline{G_i}=(V,\overline{E_i})$} with a partial
      order \boldmath{$P$} contained in \boldmath{$\overline{E_i}$},
      such that there is no transitive orientation of
      \boldmath{$\overline{G_i}$} that extends \boldmath{$P$}.}
    \label{fig:path}
  \end{center}
\end{figure}

Now consider a situation where we need to satisfy a partial order
$P=(V,A_P)$ of precedence constraints in the time dimension. It
follows that each arc $a=(u,w)\in A_P$ in this partial order forces
the corresponding undirected edge $e=\{u,w\}$ to be excluded from
$E_i$. Thus, we can simply initialize our algorithm for constructing
packing classes by fixing all undirected edges corresponding to $A_P$
to be contained in $\overline{E_i}$. After running the original
algorithm, we may get additional comparability edges. As the example
in Figure~\ref{fig:path} shows, this causes an additional problem:
Even if we know that the graph $\overline{G_i}$ has a transitive
orientation, and all arcs $a=(u,w)$ of the precedence order $(V,A_P)$
are contained in $\overline{E_i}$ as $e=\{u,w\}$, it is not clear that
there is a transitive orientation that contains all arcs of $A_P$.

\subsection{Extending Partial Suborders}
\label{sec:find-feas-trans}

Consider a comparability graph $\overline{G}$ that is the complement
of an interval graph $G$. The problem {\bf TOP} of deciding whether
$\overline{G}$ has a transitive orientation that extends a given
partial order $P$ has been studied in the context of scheduling.
Korte and M{\"o}hring \cite{KM} give a linear-time algorithm for
determining a solution, or deciding that none exists.  Their approach
is based on a very special data structure called \emph{modified
  PQ-trees}.

In principle it is possible to solve higher-dimensional packing
problems with precedence constraints by adding this algorithm as a
black box to test the leaves of our search tree for packing classes:
In case of failure, backtrack in the tree.  However, the resulting
method cannot be expected to be reasonably efficient: During the
course of our tree search, we are not dealing with one fixed
comparability graph, but only build it while exploring the search
tree.  This means that we have to expect spending a considerable
amount of time testing similar leaves in the search tree, i.e.,
comparability graphs that share most of their graph structure.  It may
be that already a very small part of this structure that is fixed very
``high'' in the search tree constitutes an obstruction that prevents a
feasible orientation of all graphs constructed below it.  So a
``deep'' search may take a long time to get rid of this obstruction.
This makes it desirable to use more structural properties of
comparability graphs and their orientations to make use of
obstructions already ``high'' in the search tree.

\subsection{Implied Orientations}
\label{sec:implied-orientations}

As in the basic packing class approach, we consider the component
graphs $G_i$ and their complements, the comparability graphs
$\overline{G_i}$.  This means that we continue to have three basic
states for any edge:
\begin{enumerate}[1:]
\item edges that have been fixed to be in $E_i$, i.e., \emph{component
    edges};
\item edges that have been fixed to be in $\overline{E_i}$, i.e.,
  \emph{comparability edges};
\item \emph{unassigned edges}.
\end{enumerate}

In order to deal with precedence constraints, we also consider
orientations of the comparability edges.  This means that during the
course of our tree search, we can have three different possible states
for each comparability edge:
\begin{enumerate}[2a:]
\item one possible orientation;
\item the opposite possible orientation;
\item no assigned orientation.
\end{enumerate}

\begin{figure}[h]
  \begin{center}
  \psfrag{(I)}{(I)}
  \psfrag{(II)}{(II)}
  \psfrag{(III)}{(III)}
  \psfrag{(I')}{(I')}
  \psfrag{(II')}{(II')}
  \psfrag{(III')}{(III')}
  \psfrag{(D1)}{(D1)}
  \psfrag{(D2)}{(D2)}
  \psfrag{v-1}{$v_1$}
  \psfrag{v-2}{$v_2$}
  \psfrag{v-3}{$v_3$}
  \psfrag{Eb}{$\overline{E}$}
  \psfrag{E}{$E$}
  \psfrag{(comparability edges)}{\textsl{\scriptsize (comparability edges)}}
  \psfrag{(component edges)}{\textsl{\scriptsize (component edges)}}
  \psfrag{unassigned or}{\textsl{\scriptsize (unassigned or}}
  \psfrag{comparability edges}{\textsl{\scriptsize comparability edges)}}
  \includegraphics[width=13cm]{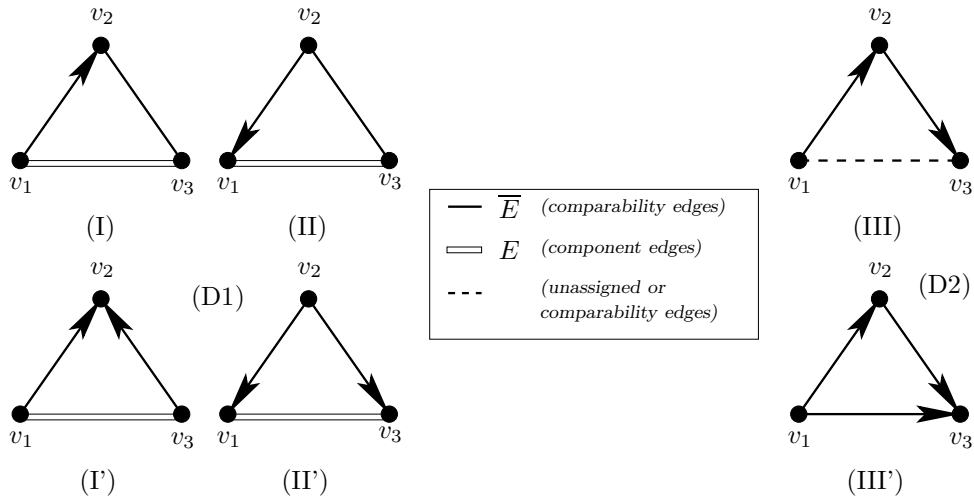}
  \caption{Implications for edges and their
    orientations: Above are $P_3$ implications (D1, left) and
    transitivity implications (D2, right); below the forced
    orientations of edges.}
  \label{fig:implication}
  \end{center}
\end{figure}

A stepping stone for this approach arises from considering the
following two configurations; see Figure~\ref{fig:implication}.
The first configuration (shown in the left part of the figure)
consists of the two comparability edges $\{v_1,v_2\}$, $\{v_2,v_3\}$
$\in \overline{E_i}$, such that the third edge $\{v_1,v_3\}$ has been
fixed to be an edge in the component graph $E_i$.  Now any orientation
of just one of the comparability edges forces the orientation of the
other comparability edge.  {In Figure~\ref{fig:implication}
  the oriented edge in (I) forces the orientation of the second edge
  as shown in (I'), similarly for (II) and (II').  Because this
  configuration corresponds to an partially oriented induced path on
  three vertices, a $P_3$ in $\overline{G_i}$, we call this
  arrangement a \emph{$P_3$ implication}.}

The second configuration (shown in the right part of the figure)
consists of two directed comparability edges $(v_1,v_2),(v_2,v_3)$. In
this case we know that edge $\{v_1,v_3\}$ must also be a comparability
edge, with an orientation of $(v_1,v_3)$. Because this configuration
arises directly from transitivity in $\overline{G_i}$,
  we call this arrangement a \emph{transitivity implication}.

Clearly, any implication arising from one of the above configurations
can induce further implications.

In particular, when considering only sequences of $P_3$ implications,
we get a partition of comparability edges into \emph{$P_3$ implication
  classes} that will be used in more detail in
Section~\ref{sec:impl-class-modul}.  Two comparability edges are in
the same $P_3$ implication class, iff there is a sequence of $P_3$
implications, such that orienting one edge forces the orientation of
the other edge.  It is not hard to see that the $P_3$
  implication classes form a partition of the
comparability edges, because we are dealing with an equivalence
relation.  For an example, consider the arrangement in
Figure~\ref{fig:path}.  Here, all three comparability edges
$\{v_1,v_2\}$, $\{v_2,v_3\}$, and $\{v_3,v_4\}$ are in the same $P_3$
implication class. Now the orientation of $(v_1,v_2)$ implies the
orientation $(v_3,v_2)$, which in turn implies the orientation
$(v_3,v_4)$, contradicting the orientation of $\{v_3,v_4\}$ in the
given partial order $P$.

We call a violation of a $P_3$ implication a \emph{$P_3$ conflict}.

\begin{figure}[htbp]
\begin{center}
  \includegraphics[width=.95\linewidth]{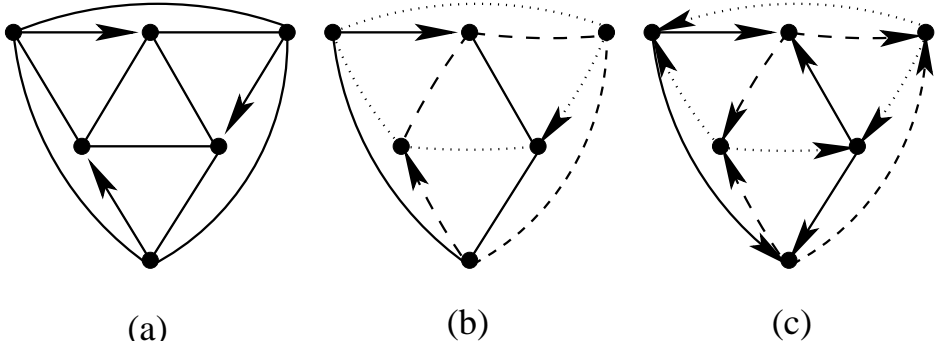}
  \caption{(a) A graph $\overline{G_i}$ with a partial order
    formed by three directed edges; (b) there are three $P_3$
    implication classes that each have one directed arc; (c) carrying
    out $P_3$ implications creates directed cycles, i.e., transitivity
    conflicts.}
  \label{fig:kreis}
\end{center}
\end{figure}

As the example in Figure~\ref{fig:kreis} shows, only excluding $P_3$
conflicts when recursively carrying out $P_3$ implications does not
suffice to guarantee the existence of a feasible orientation: Working
through the queue of $P_3$ implications, we end up with a directed
cycle, which violates a transitivity implication.

\bigskip

We call a violation of a transitivity implication a \emph{transitivity
  conflict}.

\medskip
Summarizing, we have the following necessary conditions for the
existence of a transitive orientation that extends a given partial
order $P$:

\begin{enumerate}[\bf D1:]
\item Any $P_3$ implication can be carried out without a conflict.
\item Any transitivity implication can be carried out without a
  conflict.
\end{enumerate}

These necessary conditions are also sufficient:

\medskip
\begin{theorem}\label{th:class}
  Let $P=(V,<)$ be a partial order with arc set $A_P$ that is
  contained in the edge set $E$ of a given comparability graph
  $G=(V,E)$. $A_P$ can be extended to a transitive orientation of $G$,
  iff all arising $P_3$ implications and transitivity implications can
  be carried out without creating a $P_{3}$ conflict or a transitivity
  conflict.
\end{theorem}

\medskip
A full proof and further mathematical details are described in the
following Section~\ref{sec:impl-class-modul}.  This extends
previous work by Gallai \cite{Gallai}, who extensively studied
implication classes of comparability graphs.  See Kelly~\cite{Kelly},
M{\"o}hring~\cite{RHM} for helpful
  surveys on this topic, and
Kr{\"a}mer~\cite{Kraemer} for an application in scheduling theory.

\section{Extending Partial Orientations}
\label{sec:impl-class-modul}

\paragraph{Modular decomposition}

The concept of \emph{modular decomposition} of a graph was first
introduced by Gallai \cite{Gallai} for studying comparability graphs.
This powerful decomposition scheme has a variety of applications in
algorithmic graph theory; for further material on this concept and
its application the interested reader is referred to
\cite{Kelly,Mohring86}.

\begin{figure}[htbp]
  \begin{center}
  \includegraphics[width=6cm]{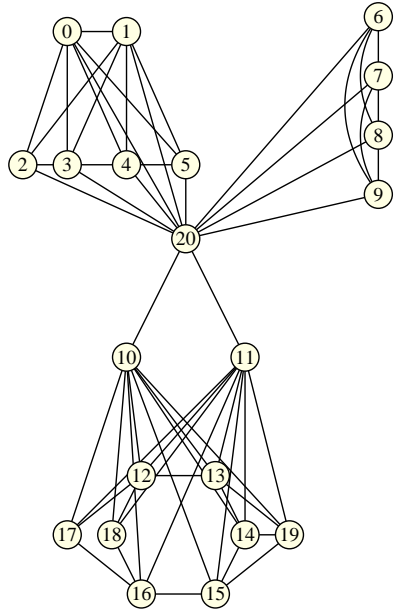}  
  \caption{An example graph $G$.}
  \label{fig:mod-graph}
  \end{center}
\end{figure}

A \emph{module} of a graph $G=(V,E)$ is a vertex set $M \subseteq V$
such that each vertex $v \in V\setminus M$ is either adjacent to all
vertices or to no vertex of $M$ in $G$. (Intuitively speaking, all
vertices of a module ``look the same'' to the other vertices of the
graph.)  A module is called trivial if $|M| \leq 1$ or $M = V$.  A
graph $G$ is called \emph{prime} if it contains only trivial modules.
Using the concept of modules one can define a decomposition scheme for
general graphs by decomposing it recursively into subsets, each of
which is a module of $G$, stopping when all sets are singletons.
{First of all, observe that every connected component
  of a given graph $G$ forms a module.  It is not hard to see that
  also every co-connected component of $G$ is a module.  If both $G$
  and its complement are connected then the decomposition needs a
  further idea.  Consider the graph in Figure~\ref{fig:mod-graph}.
  Obviously it is connected and co-connected and has a huge number of
  non-trivial modules.  However, if one identifies the \emph{maximal
    proper submodules} of $G$, i.e., those modules $M$ that are
  inclusion-maximal modules of $G$ with $M \not= V$, then one obtains
  a partition of the vertex set.  The corresponding modules of the
  example $G$ are $M_1 = \{0$, $1$, $2$, $3$, $4$, $5$, $6$, $7$, $8$,
  $9\}$, $M_2 = \{ 20 \}$, $M_3 = \{ 10, 11\}$, $M_4 = \{ 12, 13, 14,
  15, 16, 17, 18$, $19\}$.}

Gallai~\cite{Gallai} showed that any graph $G$ has a particular
decomposition (the so-called \emph{canonical decomposition}) of its
vertex set into a set of modules with a variety of nice properties.
He observed that any graph $G$ is either of \emph{parallel type},
i.e., $G$ is not connected; or $G$ is of \emph{series type}, i.e.,
$\overline{G}$ is not connected, or $G$ is of \emph{prime type}, i.e.,
$G$ and $\overline{G}$ are connected.  In the first case the canonical
decomposition is defined by the set of connected components; in the
second case the canonical decomposition is given by the connected
components of $\overline{G}$; finally, for prime-type graphs, the
canonical decomposition is given by decomposing $G$ into its maximal
proper submodules.  Gallai also showed that this decomposition is
unique.

\begin{figure}[htbp]
    \begin{center}
      \includegraphics[height=7cm]{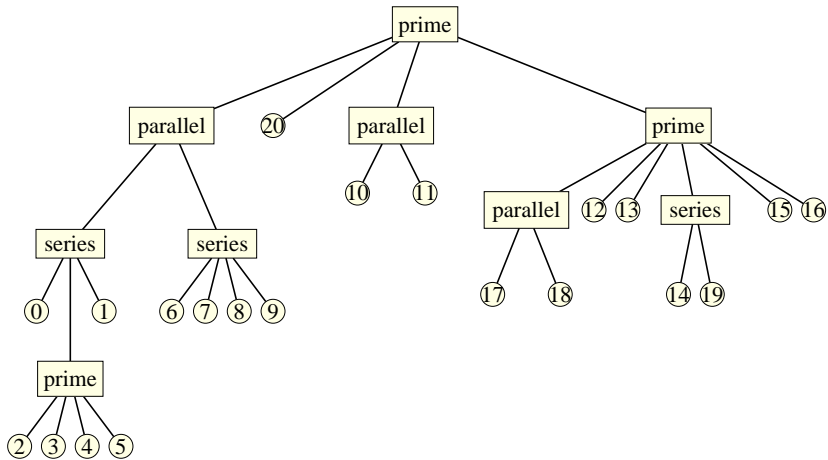}
      \caption{A modular decomposition tree for the graph $G$
        shown in Figure~\ref{fig:mod-graph}.}\label{fig:mod-dec-ex}
    \end{center}
\end{figure}

This recursive decomposition defines a \emph{decomposition tree}
$T(G)$ for a given graph $G$ in a very natural way: Create a root
vertex of $T(G)$ for the trivial module $G$ itself.  Label it series,
parallel, or prime, depending on the type of $G$.  For each
non-singleton module of the canonical decomposition of $G$ create a
tree vertex, labeled as series-, parallel-, or prime-type node,
depending on the type of the module, and make it a child of the vertex
corresponding to $G$; for each singleton module add a tree-vertex
labeled with the corresponding singleton.  Now proceed recursively for
each subgraph corresponding to a non-trivial module in the
decomposition tree, until all leaves of the tree are labeled with
singletons.  Consequently, the leaves of the tree correspond to the
vertices of the graph, while all internal vertices correspond to
non-trivial modules of the canonical decomposition of the
corresponding parent vertex in $T(G)$.  See
Figure~\ref{fig:mod-dec-ex} for the decomposition tree of our example.

The \emph{decomposition graph} $G^{\#}$ of a graph $G$ is the quotient
of $G$ by the canonical decomposition into the set of modules $\{A_1,
\dots, A_q\}$, i.e., $V(G^{\#}) = \{A_1, \dots, A_q\}$, and distinct
vertices $A_i$ and $A_j$ are joined by an edge in $G^{\#}$ iff there
is an $A_i A_j$-edge in $G$.  In the following we will look at the
decomposition graphs corresponding to internal vertices of $T(G)$ and
refer to them as the decomposition graphs of $T$.

In our example, the decomposition graph $G^{\#}$ of $G$, i.e., to the
root node of $T(G)$, is a path on four vertices, given by $$G^{\#} =
(\{M_1, M_2, M_3, M_4 \}, \{\{M_1,M_2\}, \{M_2, M_3\}, \{M_3,M_4\}
\}).$$

\medskip{}

\paragraph{Modular decomposition and transitive orientations}

An important property of the modular decomposition is its close
relationship to the concept of $P_3$ implication classes.  Gallai
observed the following properties of $P_3$ implication classes with
respect to the modular decomposition:

\medskip
\begin{proposition}[Gallai~\cite{Gallai}]\label{obs:modul-decomp-1}
  Let $G=(V,E)$ be an undirected graph.
\begin{enumerate}[1)]
\item\label{item:5} If $G$ is not connected and $G_1, \dots, G_q$ ($q
  \geq 2$) are the components of $G$, then the $P_3$ implication
  classes of $G_1, \dots, G_q$ are exactly the $P_3$ implication
  classes of $G$.
\item\label{item:6} If $\overline{G}$ is not connected (so that $G$ is
  connected), $\overline{G_1}, \dots, \overline{G_q}$ ($q \geq 2$) are
  the components of $\overline{G}$, and $A_i = V(G_i)$, then $A_i$ and
  $A_j$ are completely connected to each other whenever $1 \leq i < j
  \leq q$.  Moreover, for all such $i$ and $j$, the set of $A_i
  A_j$-edges form an $P_3$ implication class $E_{i j}$ of $G$.  The
  $P_3$ implication classes of $G$ that are distinct from any $E_{i j}$
  are exactly the $P_3$ implication classes of the graphs $G_i =
  G[A_i]$ ($i = 1, \dots, q$).
\item\label{item:7} If $G$ and $\overline{G}$ are both connected and
  have more than one vertex, and the canonical decomposition of $G$ is
  given by $\{A_1, \dots, A_q\}$, then we have
  \begin{enumerate}[a)]
  \item\label{item:1} If there is one edge between $A_i$ and $A_j$ ($1
    \leq i < j \leq q$), then all edges between $A_i$ and
    $A_j$ are in $G$.
  \item\label{item:2} The set of all edges of $G$ that join
      different $A_i$s forms a single
    $P_3$ implication class $C$ of $G$.  Every vertex of $G$ is
    incident with some edge of $C$, (i.e., $V(C) = V(G)$).
  \item\label{item:3} The $P_3$ implication classes of $G$ that are
    distinct from $C$ are exactly the $P_3$ implication classes of the
    graphs $G_i = G[A_i]$ ($1 \leq i \leq q$).
  \end{enumerate}
\end{enumerate}
\end{proposition}
  
\medskip
This strong relationship between $P_3$ implication classes and the
modules in the canonical decomposition of a given graph
is a powerful tool for studying
graphs having a transitive orientation.  Note that the fastest known
algorithms for recognizing comparability graphs 
  make extensively use of this relationship.
Gallai used the above properties (among others) for proving the
following theorem.

\medskip
\begin{theorem}[Gallai~\cite{Gallai}]\label{thm:gallai-1.9}
  Let $G$ be a non-empty graph, let $T = T(G)$ be the tree
  decomposition of $G$, and let $H$ be a vertex set corresponding to a
  node of $T$.
  \begin{enumerate}[1)]
  \item\label{item:8} If $G$ is transitively oriented, and $A$ and $B$
    are descendents of $H$ in $T$, then every $A, B$-edge of $G$ is
    oriented in the same direction (to or from $A$).  Therefore,
    $H^{\#}$ receives an induced transitive orientation.
  \item\label{item:9} Conversely, assuming that $H^{\#}$ is
    transitively orientable for each $H \in T$, one can choose an
    arbitrary transitive orientation of each $H^{\#}$ and induce a
    transitive orientation of $G$ by orienting all $A, B$-edges (for
    $A$ and $B$ descendents of $H$ in $T$) in the same direction that
    $\{A, B\}$ is oriented in $H^{\#}$.
  \end{enumerate}
\end{theorem}

\medskip
It is straightforward to draw the following helpful corollaries from
this theorem:

\medskip
\begin{corollary}
  A graph $G$ is a comparability graph if and only if every
  decomposition graph in the tree decomposition of $G$ is a
  comparability graph.
\end{corollary}

\medskip
\begin{corollary}\label{thm:der-hammer}
  Let $G$ be a comparability graph and $T$ its tree decomposition.
  Assigning to each of the decomposition graphs of $T$ a transitive
  orientation independently results in a transitive orientation of
  $G$.
\end{corollary}

\medskip
Furthermore, if only a partial orientation of $G$ is given and we are
interested in extending this orientation to a transitive orientation
of $G$, we can formulate the following lemma.

\medskip
\begin{lemma}\label{cor:extend-partial}
  Let $G$ be a comparability graph and $T$ its tree decomposition.
  Furthermore, let $P$ be a partial orientation of $G$, assigning
  orientations to some, but not all $P_3$ implication classes of $G$.
  $P$ is extendible to a transitive orientation of $G$ if and only if
  for each decomposition graph $H^{\#}$ of $T$ the orientation induced
  on $H^{\#}$ by $P$ is extendible to a transitive orientation on
  $H^{\#}$.
\end{lemma}

\medskip
\textbf{Proof:} Follows immediately from Theorem~\ref{thm:gallai-1.9}
(\ref{item:9}).  \hfill$\Box$

\medskip

Now we are ready to prove Theorem~\ref{th:class}:
Conditions {\bf D1} and {\bf D2} are also sufficient.

\paragraph{Proof of Theorem~\ref{th:class}:}
Suppose there is a transitive orientation $F$ of $G$ that contains
$P$.  Because $F$ is a transitive orientation, all arcs implied by
$P_3$ or transitivity implications are contained in $F$.  Furthermore,
there cannot be any $P_3$ or transitivity conflict in $F$, again
because $F$ is a transitive orientation.  Thus $F$ shows that all
arising $P_3$ and transitivity implications can be carried out without
creating a $P_3$ or transitivity conflict.
    
Suppose now that \textbf{D1} and \textbf{D2} are satisfied, i.e.,
there is a directed graph $F$ consisting of all arcs of $P$ together
with all orientations of edges of $G$ that are implied by a sequence
of $P_3$ and transitivity implications of arcs of $P$.  In other words,
$F$ contains all arcs that are forced by $P_3$ or transitivity implications
together with all their implied arcs; i.e., all arcs that are forced
by arcs of $F$ are also contained in $F$.  We show that $F$ can be
extended to a transitive orientation of $G$.
  
First observe that, by assumption, there cannot be a $P_3$ or
transitivity conflict in $F$.  In particular, $F$ is an orientation of
edges of $G$ and for each $P_3$ implication class $C$ of $G$ that has
at least one edge that is oriented in $F$, all edges of $C$ are
oriented in $F$ and this orientation is conflict-free.  By
Corollary~\ref{thm:der-hammer},
every single conflict-free oriented $P_3$ implication class of $G$ by
itself is extendible to a transitive orientation of $G$.
  
Now let $T$ be the decomposition tree of $G$ and consider the
decomposition graphs corresponding to $T$.  By the above observation,
the orientation of an $P_3$ implication class $C$ in $F$ implies an
orientation of the edge(s) corresponding to this $P_3$ implication
class in the decomposition graphs of $T$.  More precisely, by
Observation~\ref{obs:modul-decomp-1} (\ref{item:6}), for every 
series-type node $H$ of $T$ each edge $e = \{A B\}$ of $H^{\#}$ corresponds
exactly to one $P_3$ implication class $C_e$ of $G$.  If $C_e$ is
oriented conflict-free in $F$, this orientation directly induces an
orientation of $e$ (see Theorem~\ref{thm:gallai-1.9}).  For a prime-type 
node $H$ the set of edges joining different $A_i$s forms exactly
one $P_3$ implication class $C_E$ of $G$ (see
Observation~\ref{obs:modul-decomp-1} (\ref{item:7})).  Again, if $C_E$
is oriented conflict-free in $F$, this orientation immediately implies
an orientation on $H^{\#}$.
  
All we have to show now is that for each decomposition graph $H^{\#}$
of $T$, the partial orientation implied by $F$ can be extended to a
transitive orientation of $H^{\#}$.  Then, by
Corollary~\ref{thm:der-hammer}, the implied orientation of $G$ is
transitive.
  
By Corollary~\ref{thm:der-hammer}, a \emph{parallel-type node} of $T$
cannot create a contradiction to transitivity---it does not contain
any edges.

Also a \emph{prime-type node} of $T$ cannot create a contradiction:
All of its edges are contained in only one $P_3$ implication class and,
because all $P_3$ implication classes of $G$ contained in $F$ are
oriented conflict-free, the corresponding orientation induced by $F$
on this single $P_3$ implication class has to be transitive.
  
This leaves the case of \emph{series-type nodes}.
  Suppose there is a series-type node $H$ of $T$ with
decomposition graph $H^{\#}$, for which the partial orientation implied by
$F$ cannot be extended to a transitive orientation of $H^{\#}$.  Then
we claim that this partial orientation has to be cyclic: By definition
for each series-type node $H$ of $T$ the decomposition graph $H^{\#}$
is a complete graph and every acyclic partial orientation of a
complete graph can be extended to a transitive orientation of this
complete graph by taking any topological ordering of the vertices that
agrees with the partial orientation.  Hence, the partial orientation
on $H^{\#}$ has to contain a directed cycle.
  
However, by the definition of $T$ and the implied orientation of
$H^{\#}$ by $F$, a directed cycle in $H^{\#}$ immediately implies a
cyclically oriented cycle in $F$.  Furthermore, with every consecutive
pair of oriented edges $(x, y)$, $(y, z)$ of this cycle also the
oriented edge $(x, z)$ (which is implied by transitivity) has to be
contained in $F$.  Iterating this argument results in an cyclically
oriented triangle in $F$, which is a transitivity conflict. This
contradicts our assumption that there are no transitivity conflicts.
\hfill$\Box$

\section{Computational Experiments}\label{sec:exp}

\subsection{Solving Problems with Precedence Constraints}
\label{sec:solv-probl-with}

We start by fixing for all arcs $(u,v)\in A_P$
  the edge $\{u,v\}$ as an edge in the comparability
graph $\overline{G_i}$, and we also fix its orientation to be $(u,v)$.
In addition to the tests for enforcing the conditions for unoriented
packing classes ({\bf C1}, {\bf C2}, {\bf C3}), we employ the implications suggested by
conditions {\bf D1} and {\bf D2}.  For this purpose we check directed edges in
$\overline{G_i}$ for being part of a triangle that gives rise to
either implication.  Any newly oriented edge in $\overline{G_i}$ gets
added to a queue of unprocessed edges. Like for packing classes, we
can again get cascades of fixed edge orientations.  If we get an
orientation conflict or a cycle conflict, we can abandon the search on
this tree node.  The correctness of the overall algorithm follows from
Theorem~\ref{th:class}; in particular, the theorem guarantees that we
can carry out implications in an arbitrary order.  In the following we
present our results for different types of instances: The video-codec
benchmark described in Section~\ref{video} arises from an actual
application to FPGAs.  In Section~\ref{math} we give a number of
results arising from different geometric packing problems.

Our code was implemented in C++ and was run
on a SUN Ultra 10 with 333 MHz.

The first example is a numerical method for solving a differential
equation (DE) with 11 nodes.  The node operations are either
multiplications or ALU-type operations.  In a second example, a
video-codec using the H.261 norm is optimized.  These examples are
meant to demonstrate the general applicability of our method for
practical problems; given other problem instances, or additional
constraints, we can easily adapt our algorithm.

\subsection{DE Benchmark}
\label{sec:de-benchmark}

Let the module library contain two hardware modules (box types): an
array-multiplier and a module of type ALU that realizes all other node
operations (comparison, addition, subtraction).  For a word-length of
n=16 bits, we assume a module geometry of 16 x 1 cells for the ALU
module, and of 16 x 16 cells for the multiplier.  Furthermore, the
execution time of an ALU node takes one clock cycle, while a
multiplication requires 2 clock cycles on our target chip.

The dependency graph is shown in Fig.~\ref{fig:pdedep}.  First, we
compute the transitive closure of all data dependencies to allow our
algorithm to find contradictions to feasible packings already in the
input.

Next, we solve several instances of the BMP problem for different
values of $h_t$ reported in Table~\ref{tab1}.  Each $h_t$ listed
yields a test case for which the container size is minimized
(\emph{MinA}\,), assuming $h_x=h_y$.  Also shown is the CPU-time
needed for finding a solution.

\begin{table}[htb]
\footnotesize
  \caption{Computational results for optimizing reconfigurations for
  the DE benchmark.}\label{tab1}
  \begin{center}
  \begin{tabular}{c c c c c}
  \toprule
  test & \multicolumn{3}{c}{container sizes} &  \\ 
  \cmidrule{2-4}
        & $h_t$ & $h_x$ & $h_y$ & CPU-time \\
  \midrule
  1     &   6 &   32  & 32  &  55.76 s \\
  2     &  13 &   17  & 17  &  0.04 s \\
  3     &  14 &   16  & 16  &  0.03 s \\
  \bottomrule
 \end{tabular}
 \end{center}
\end{table}

The reported optimization times were measured as the CPU-times on a
SUN-Ultra 10 with 333 MHz.
   
For the DE benchmark, it turns out that a chip of 32 x 32 freely
programmable cells is necessary to obtain a latency between 6 and 12
clock cycles.  As the longest path in the graph has length 6, there
does not exist any faster schedule.  For 12 and 13 cycles, a chip of
size 17 x 17 is necessary, for $h_t \geq 14$, a chip of size 16 x 16
cells is sufficient, which is the smallest chip possible to implement
the problem, as one multiplication by itself uses the full chip.

\begin{figure}[htbp]
  \begin{center}
    \includegraphics[width=5cm]{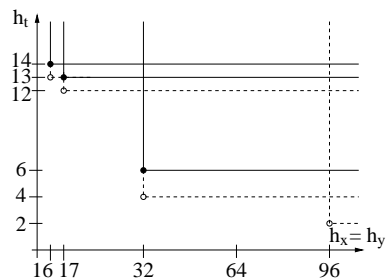}
    \caption{Pareto-optimal points for minimizing chip
      area and processing time for the DE benchmark. (a) Including
      partial order constraints (solid lines).  (b) Without
      consideration of partial order constraints (dashed lines).}
    \label{fig:depareto}
  \end{center}
\end{figure}

The SPP is solved in a similar way.  The tradeoff between area size and
necessary time is visualized in Fig.~\ref{fig:depareto}, in which the
Pareto-optimal points are shown. The figure also shows the Pareto
points for the case where no partial order needs to be satisfied
(shown dashed).

\subsection{Video-Codec Benchmark}
\label{video}

\begin{figure}[htbp]
  \begin{center}
    \includegraphics[width=.5\linewidth]{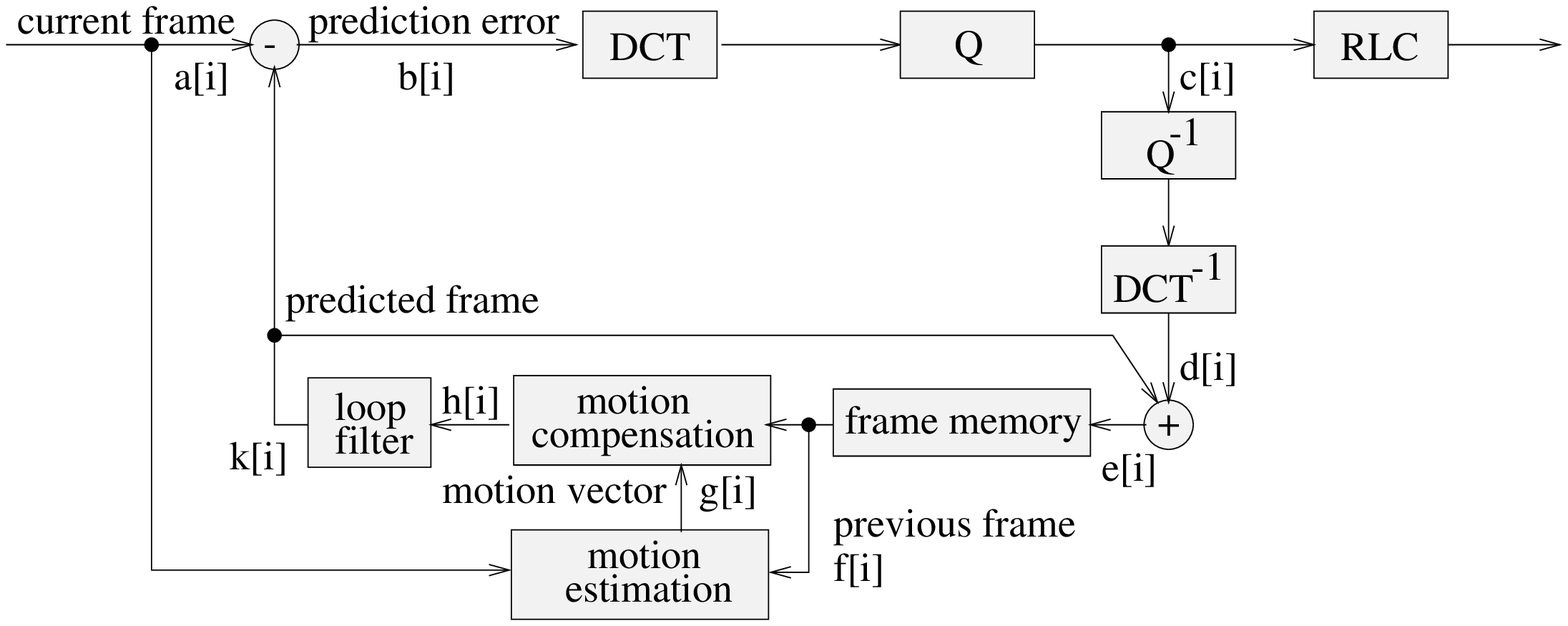}\hfill
    \caption{Block diagram of a video-codec (H.261).
      }
    \label{fig:video}
  \end{center}
\end{figure}

Figure~\ref{fig:video} shows a block diagram of the operation of a
hybrid image sequence coder/decoder that arises from the FPGA
application. The purpose of the coder is to compress video images
using the H.261 standard.  In this device, transformative and
predictive coding techniques are unified.  The compression factor can
be increased by a predictive method for motion estimates: blocks
inside a frame are predicted from blocks of previous images.

The blocks of the operational description shown in the figure
possess the granularity of more complex functions.  However, this
description contains no information corresponding to timing,
architecture, and mapping of blocks onto an architecture.
The resulting problem graph contains a subgraph for
the coder and one subgraph for the decoder.

For realizing the device we have a library of three different
modules.  One is a simple processor core with a (normalized) area
requirement of 625 units (25 x 25 cells, normalized to other modules
in order to obtain a coarser grid) called PUM, denoted by ``P'' in
Table~\ref{tab3}.  Secondly, there are two dedicated special-purpose
modules: a block matching module (BMM), ``B'' in Table~\ref{tab3})
that is used for motion estimation and requires 64 x 64 = 4096 cells;
and a module DCTM (``D'' in Table~\ref{tab3}) for computing
DCT/IDCT-computations, requiring 16 x 16 = 256 cells.  Again, the BMP
and the CSPP were considered, and the makespan was minimized for
different latency constraints.  Here there is only one Pareto-point
found, shown in Table~\ref{tab3}.

\begin{table}[htb]
\footnotesize
  \caption{Optimizing reconfigurations for the Video-Codec.}
  \label{tab3}
  \begin{center}
  \begin{tabular}{c c c c c}
  \toprule
  test & \multicolumn{3}{c}{container sizes} & \\ 
  \cmidrule{2-4}
        & $h_t$ & $h_x$ & $h_y$ & CPU-time \\
  \midrule
  1     &   59 &   64  & 64  &  24.87 s \\ 
  \bottomrule
  \end{tabular}
  \end{center}
\end{table}

\subsection{Geometric Instances}
\label{math}

We describe computational results for two types of two-dimensional
objects. See Table~\ref{tab4} for an overview. The first class of
instances was constructed from a particularly difficult random
instance of the 2-dimensional knapsack problem (see \cite{ESA}).  Results are
given for order constraints of increasing size. In order to give a
better idea of the computational difficulty, we give separate running
times for finding an optimal feasible solution, and for proving that
this solution is best possible.

\begin{table}[htb]
\footnotesize
  \caption{Optimal packing with order constraints.}
  \label{tab4}
  \begin{center}
  \begin{tabular}{c c c r r}
  \toprule
  instance & optimal & &  upper & lower \\ 
  & $h_t$ & $h_x$ &  bound & bound\\ 
  \midrule
  okp17-0&   169 &    100 &  7.29 s & 179 s \\ 
  okp17-1 &   172 &    100 &  6.73 s & 1102 s \\ 
  okp17-2 &   182 &    100 &  5.39 s & 330 s \\ 
  okp17-3 &   184 &    100 &  236 s & 553 s \\ 
  okp17-4 &   245 &    100 &  0.17 s & 0.01 s \\ 
  \addlinespace
  square21-no&   112 &    112 &  84.28 s & 0.01 s \\ 
  square21-mat&   117 &    112 &  15.12 s & 277 s \\ 
  square21-tri&   125 &    112 &  107 s & 571 s \\ 
  square21-2mat&   [118,120] & [118,120] &  346 s & 476 s \\ 
  \hline 
  \end{tabular}
  \end{center}
\end{table}

\begin{figure}[htbp]
\begin{center}
\hfill
\epsfig{figure=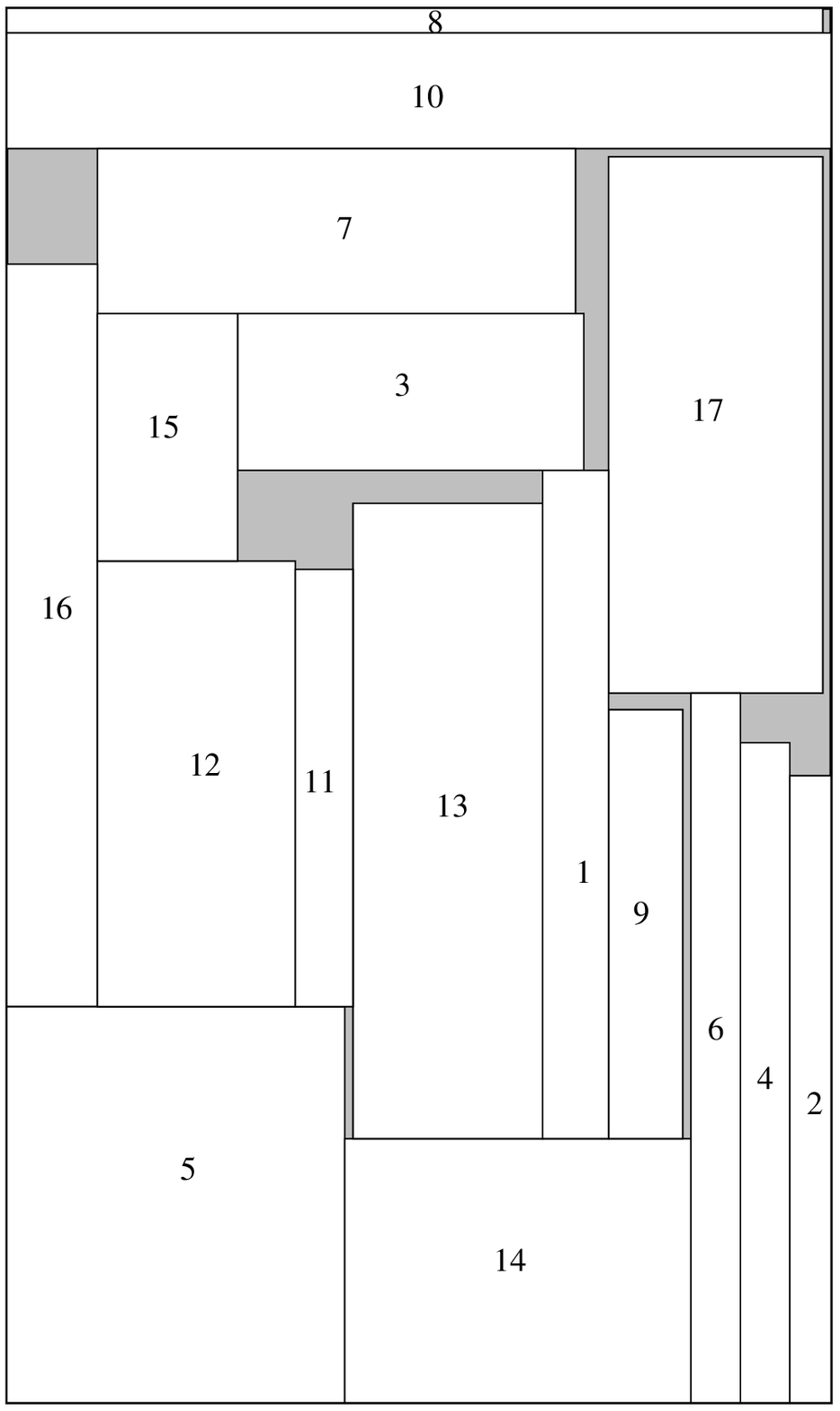,width=.32\linewidth}
\hfill
\epsfig{figure=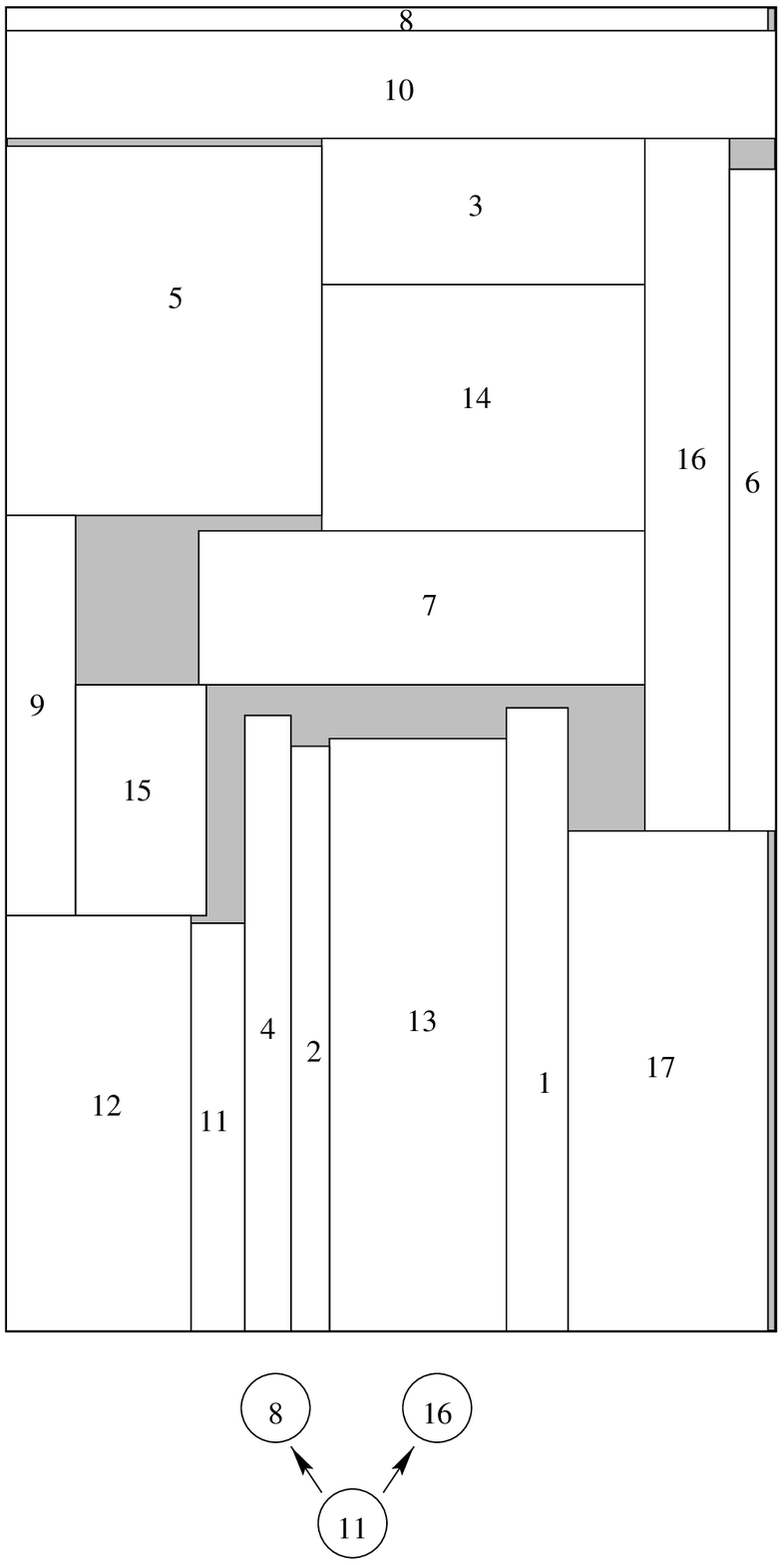,width=.32\linewidth} 
\hfill~\\
\hfill
\epsfig{figure=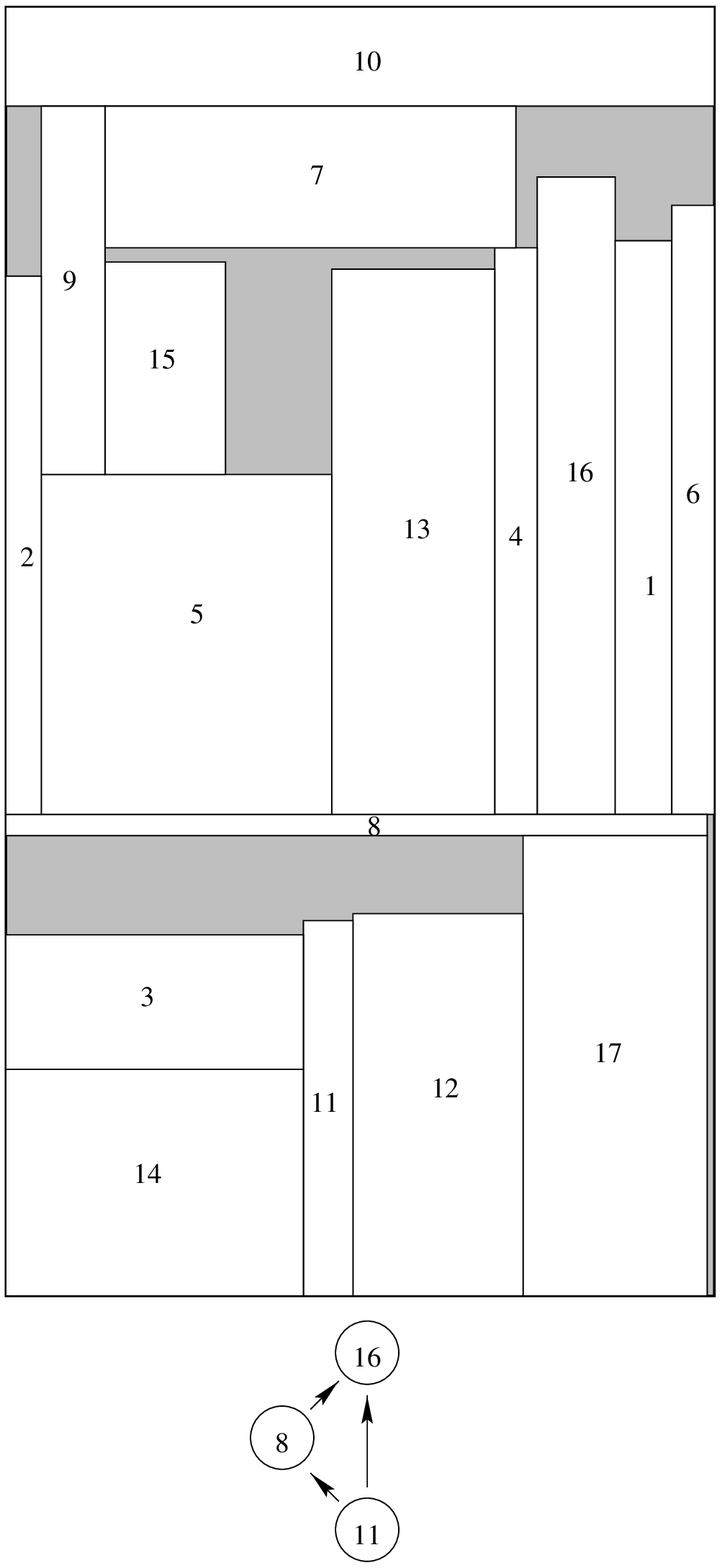,width=.32\linewidth}
\hfill
\epsfig{figure=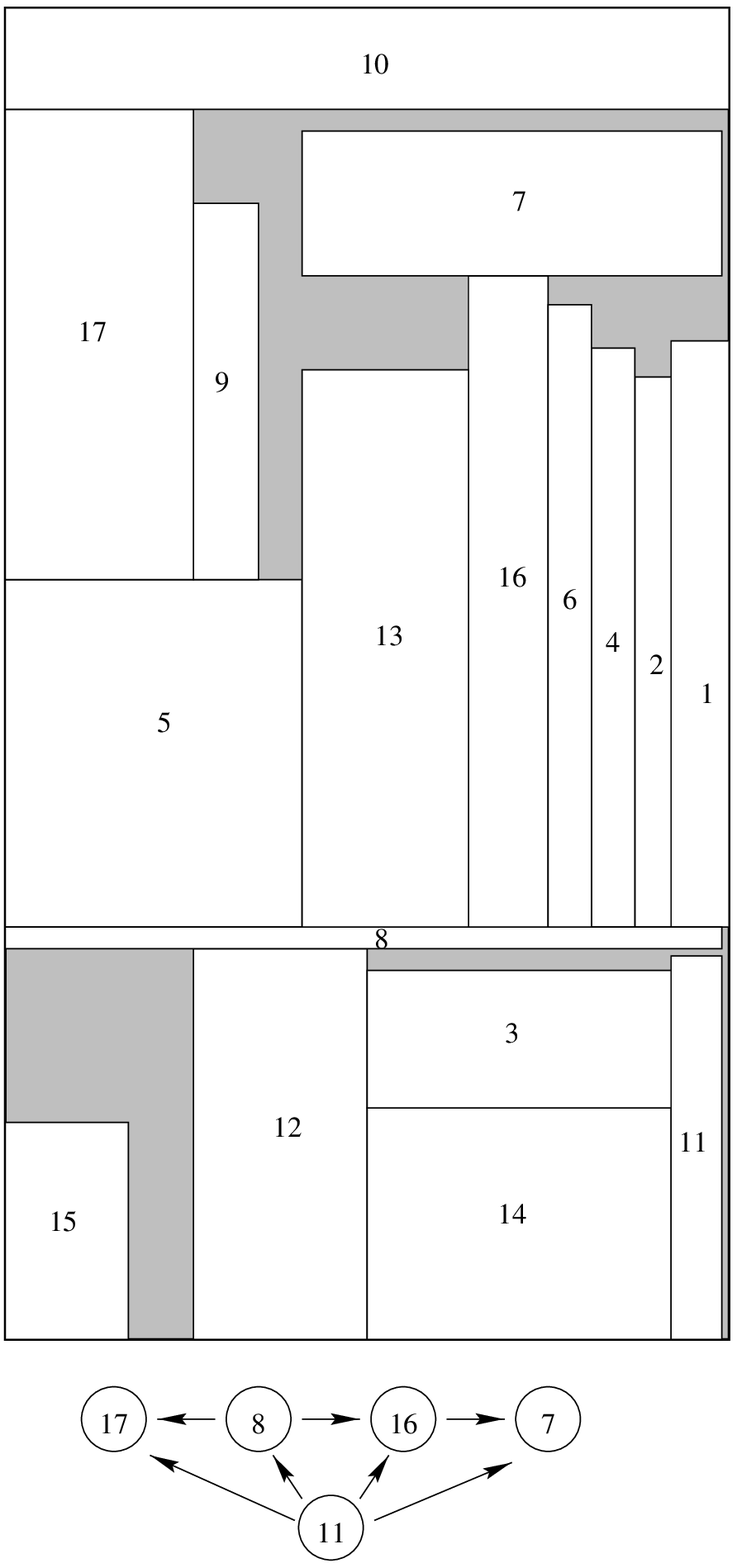,width=.32\linewidth}
\hfill~\\
\end{center}
\caption{(a) An optimal packing of okp17-0 of height 169;
  (b) an optimal packing of okp17-1 of height 172; (c) an optimal
  packing of okp17-2 of height 182; (d) An optimal packing of okp17-3
  of height 184.}
\label{fig:17}
\end{figure}

See Table~\ref{tab4} for the exact sizes of the 17 rectangles
involved, and Figure~\ref{fig:17} for the geometric layout of optimal
packings.  For easier reference, the boxes in the {\tt okp17}
instances are labeled 1-17 in the given order.

\begin{figure}[htbp]
\begin{center}
\hspace*{5mm}
\epsfig{figure=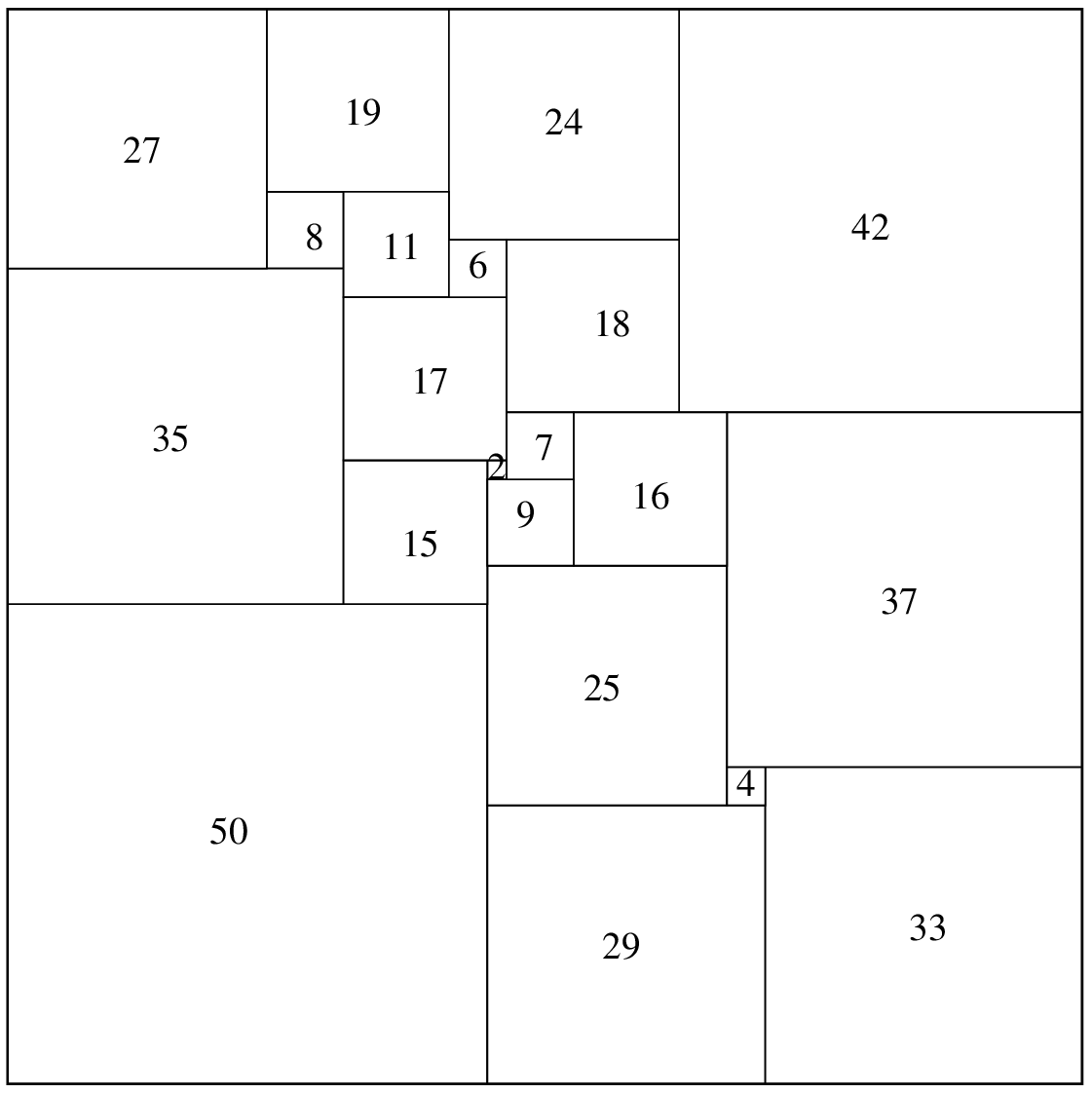,width=.45\linewidth}
\hspace*{-7mm}
\epsfig{figure=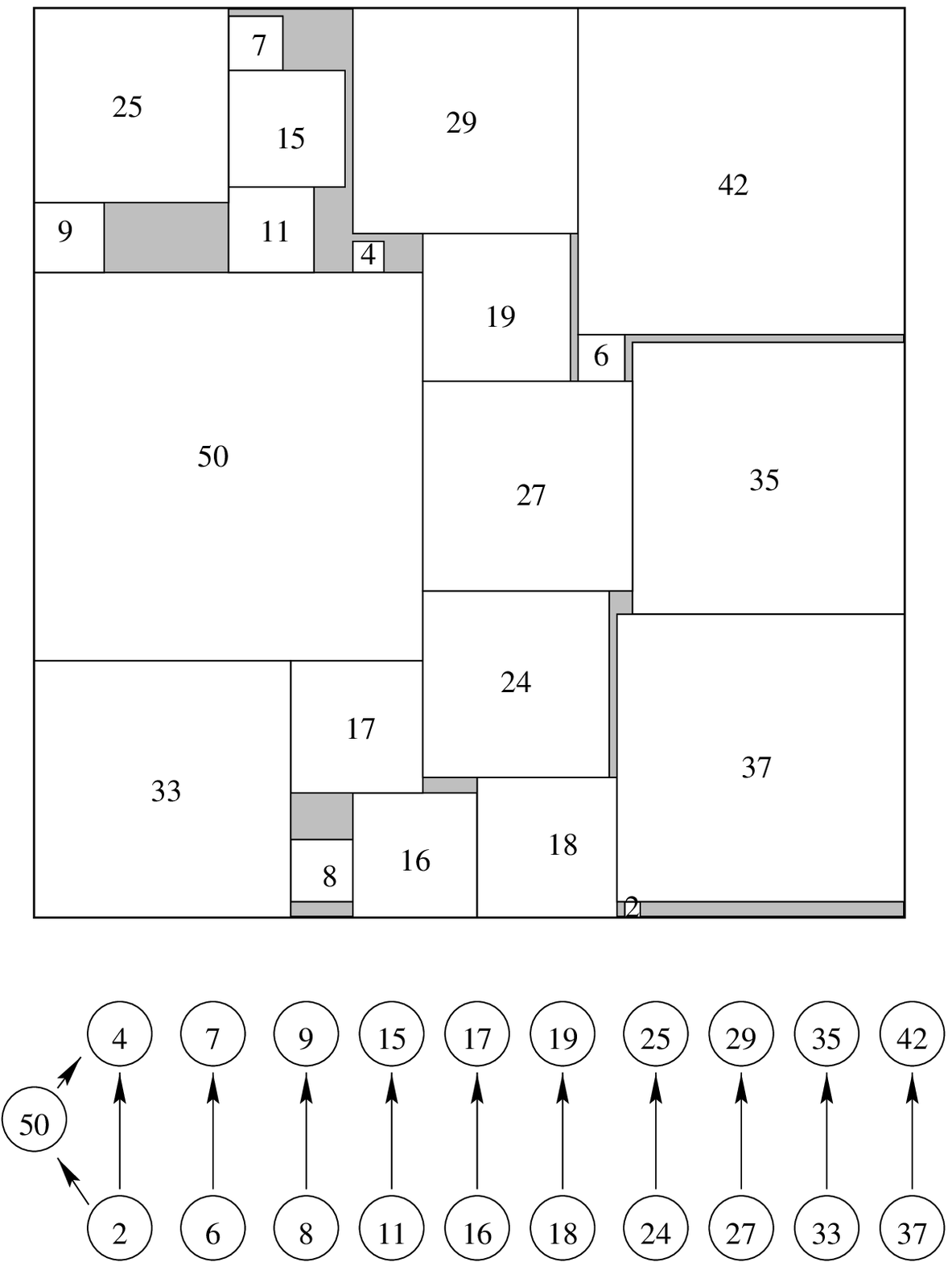,width=.45\linewidth}
\ \vspace*{5mm}
\hfill
\epsfig{figure=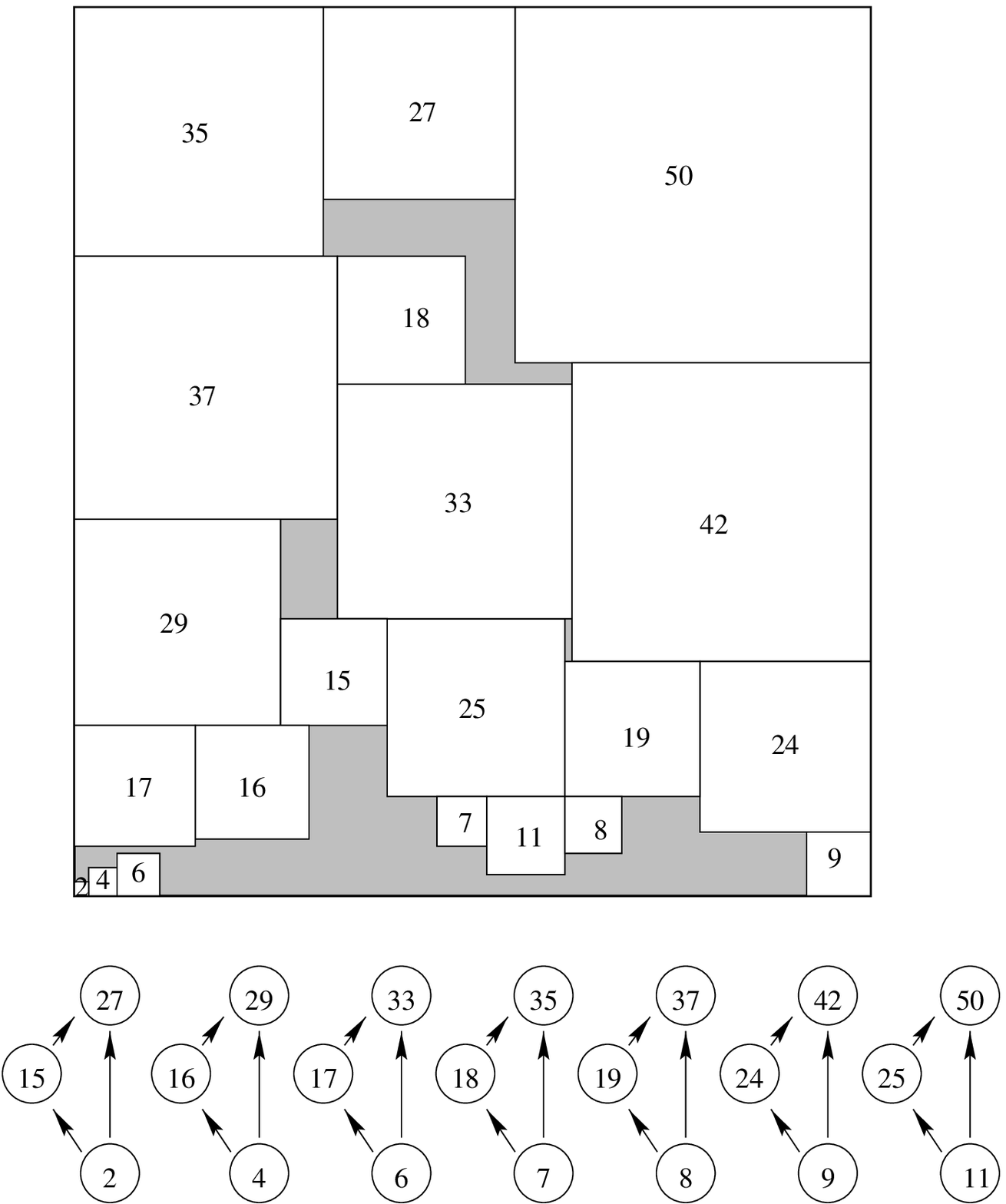,width=.45\linewidth}
\epsfig{figure=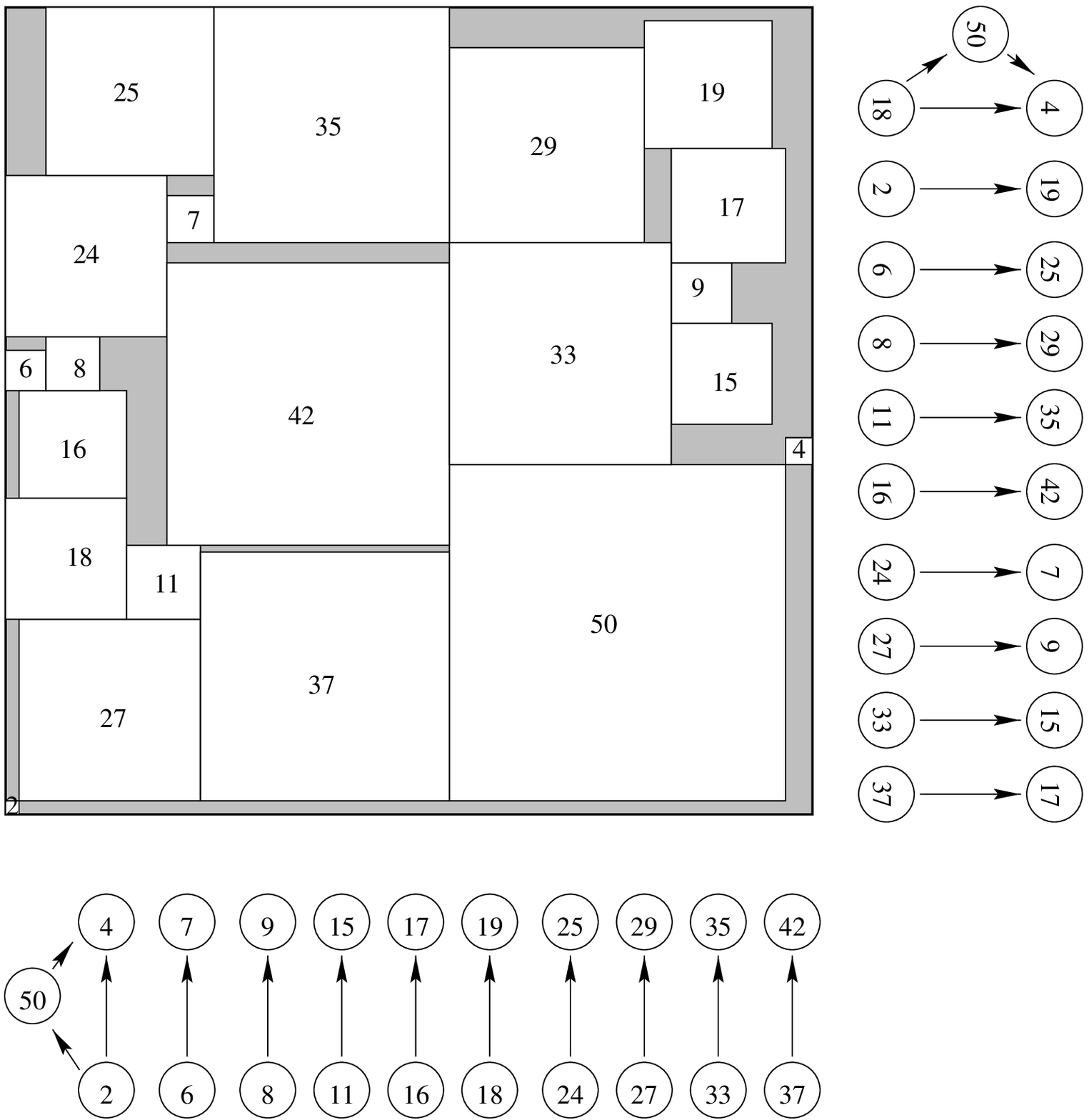,width=.45\linewidth}
\hfill
\end{center}
\caption{(a) An optimal packing of square21-0 of height 112;
    (b) an optimal packing of square21-mat of height 117; (c) an
    optimal packing of square21-tri of height 125; (d) a packing of
    square21-2mat of size 120x120.}
\label{fig:sq}
\end{figure}

The second class of instances arises from the well-known tiling of a
112x112 square by 21 squares of different sizes.  Again we have added
order constraints of various sizes.
For the instance square21-2mat (with order constraints in two
dimensions), we could not close the gap between upper and lower bound.
For this instance we report the running times for achieving the best
known bounds.  Layouts of best solutions are shown in
Figure~\ref{fig:sq}.

\begin{table}[hbtp]
\footnotesize{
  \caption{The problem instances {\tt okp17}.}
  \label{tab:okp17}
  \begin{center}
  \begin{tabular}{r l}
  \toprule
  {\tt okp17}: & base width of container = 100, number of boxes = 17 \\
  \emph{sizes} =&[(8,81),(5,76),(42,19),(6,80),(41,48),(6,86),(58,20),(99,3),(9,52),\\
  &(100,14),(7,53),(24,54),(23,77),(42,32),(17,30),(11,90),(26,65)]\\
  \addlinespace
  okp17-0: & no order constraints\\
  \addlinespace
  okp17-1: & 11$\rightarrow$8, 11$\rightarrow$16\\
  \addlinespace
  okp17-2: & 11$\rightarrow$8, 11$\rightarrow$16, 8$\rightarrow$16\\
  \addlinespace
  okp17-3: & 11$\rightarrow$8, 11$\rightarrow$16, 8$\rightarrow$16, 8$\rightarrow$17, 11$\rightarrow$7, 16$\rightarrow$7\\
  \addlinespace
  okp17-4: & 11$\rightarrow$8, 11$\rightarrow$16, 8$\rightarrow$16, 8$\rightarrow$17, 11$\rightarrow$7, 16$\rightarrow$7, 17$\rightarrow$16\\
  \bottomrule
  \end{tabular}
  \end{center}
}
\end{table}

\begin{table}[hbtp]
\footnotesize
  \caption{The problem instances {\tt square21.}}
  \label{tab:sq21}
  \begin{center}
  \begin{tabular}{r l}
  \toprule
  {\tt square21}: & base width of container = 112, number of boxes = 21 \\
  \emph{sizes} =&[(50,50),(42,42),(37,37),(35,35),(33,33),(29,29),(27,27),(25,25),\\
  &(24,24),(19,19),(18,18),(17,17),(16,16),(15,15),(11,11),(9,9),(8,8),\\
  &(7,7),(6,6),(4,4),(2,2)]\\
  \addlinespace
  square21-0: & no order constraints\\
  \addlinespace
  square21-mat: & 2$\rightarrow$4, 6$\rightarrow$7, 8$\rightarrow$9, 11$\rightarrow$15, 16$\rightarrow$17, 18$\rightarrow$19, 24$\rightarrow$25, 
  27$\rightarrow$29,\\
  &33$\rightarrow$35, 37$\rightarrow$42, 2$\rightarrow$50, 50$\rightarrow$4\\
  \addlinespace
  square21-tri: & 2$\rightarrow$15, 15$\rightarrow$17, 2$\rightarrow$27, 4$\rightarrow$16, 16$\rightarrow$29, 4$\rightarrow$29, 6$\rightarrow$17, 
  17$\rightarrow$33,\\
  & 6$\rightarrow$33, 7$\rightarrow$18, 18$\rightarrow$35, 7$\rightarrow$35, 8$\rightarrow$19, 19$\rightarrow$37, 8$\rightarrow$37, 9$\rightarrow$24,\\
  & 24$\rightarrow$42, 9$\rightarrow$42, 11$\rightarrow$25, 25$\rightarrow$50, 11$\rightarrow$50  \\
  \addlinespace
  square21-2mat: & $x$-constraints:\\
  & 2$\rightarrow$19, 6$\rightarrow$25, 8$\rightarrow$29, 11$\rightarrow$35, 16$\rightarrow$42, 18$\rightarrow$4, 24$\rightarrow$7, 
  27$\rightarrow$9,\\
  &33$\rightarrow$15, 37$\rightarrow$17, 50$\rightarrow$4, 18$\rightarrow$50\\
  & $y$-constraints:\\
  & 2$\rightarrow$4, 6$\rightarrow$7, 8$\rightarrow$9, 11$\rightarrow$15, 16$\rightarrow$17, 18$\rightarrow$19, 24$\rightarrow$25, 
  27$\rightarrow$29,\\
  & 33$\rightarrow$35, 37$\rightarrow$42, 2$\rightarrow$50, 50$\rightarrow$4\\
  \bottomrule
  \end{tabular}
  \end{center}
\end{table}

\section*{Acknowledgments}

We are extremely grateful to J{\"o}rg Schepers for letting us continue
the work with the packing code that he started as part of his thesis,
and for several helpful hints, despite of his departure to industry.
We thank Nicole Megow for helpful comments, 
Marc Uetz for a useful discussion on resource-constrained scheduling,
and an anonymous referee for a number of helpful suggestions that helped
to improve the presentation of this paper.

\newpage

\nocite{ORLIB90,HJ90,KOMO89}
\bibliography{paper}
\end{document}